# Reviewing climate change attribution in UK natural hazards and their impacts

Regan Mudhar[1,2], Dann M. Mitchell[1], Peter A. Stott[2,3], Richard A. Betts[2,3]

1 School of Geographical Sciences, University of Bristol, Bristol, UK
2 Faculty of Environment, Science and Economy, University of Exeter, Exeter, UK
3 Met Office Hadley Centre, Exeter, UK

## Executive Summary

The field of Detection and Attribution is rapidly moving beyond weather and climate, towards incorporating hazards and their impacts on natural and human systems. Here, we review the comprehensive literature base relevant for the UK ahead of the next Climate Change Risk Assessment. The current literature highlights a detectable and non-trivial influence of climate change in many UK impact sectors already – notably health, agriculture, and infrastructure. For precipitation-dominated hazards, we had medium-to-high confidence in a detected rise in event likelihood (Figure 1 shows a summary), suggesting a similar signal in related impact sectors. Sectors affected by wind-dominated hazards also displayed increased risk but with consistently low confidence. Hazards for which we had lower confidence in attribution statements typically had studies that used a variety of attribution methodologies but reached differing conclusions; for wind, this was compounded by a lack of attribution studies in general. For sectors where temperature dominated, the picture was more complex. We found that heatwaves were the most studied hazard overall, with a unanimous consensus on a strong attributable signal of human-induced climate change in their increased frequency and intensity over the last century. Though with fewer studies, the converse reduction in cold wave risk was similarly strongly attributable. The exacerbation of extreme heat events has seen an increasing impact on summer health particularly, but easing of winter health outcomes, alongside changes across various infrastructure sectors. Nevertheless, the most notable gap identified overall was in attributing climate-related impacts to human influence, with a few impact studies available for only a handful of the hazards assessed. Furthermore, just under half of the 29 hazards were not found to have any UK-relevant attribution studies, with most of the remainder having three or fewer. The majority of studies covered the UK as a whole, with England the next biggest focus, notably lagged by Scotland, then Wales, and none specifically covering Northern Ireland. All together, these remaining gaps in evidence for the attributable impact of climate change across the UK nations hinders our adaptation capability. Greater geographic granularity could benefit devolved administrations and even individual cities where city councils have decision making capabilities. This review highlights requirements for and opportunities to develop attribution science to meet the needs of the UK. Diversifying the hazards and impacts studied, in conjunction with the techniques and approaches used, will undoubtedly benefit the community.

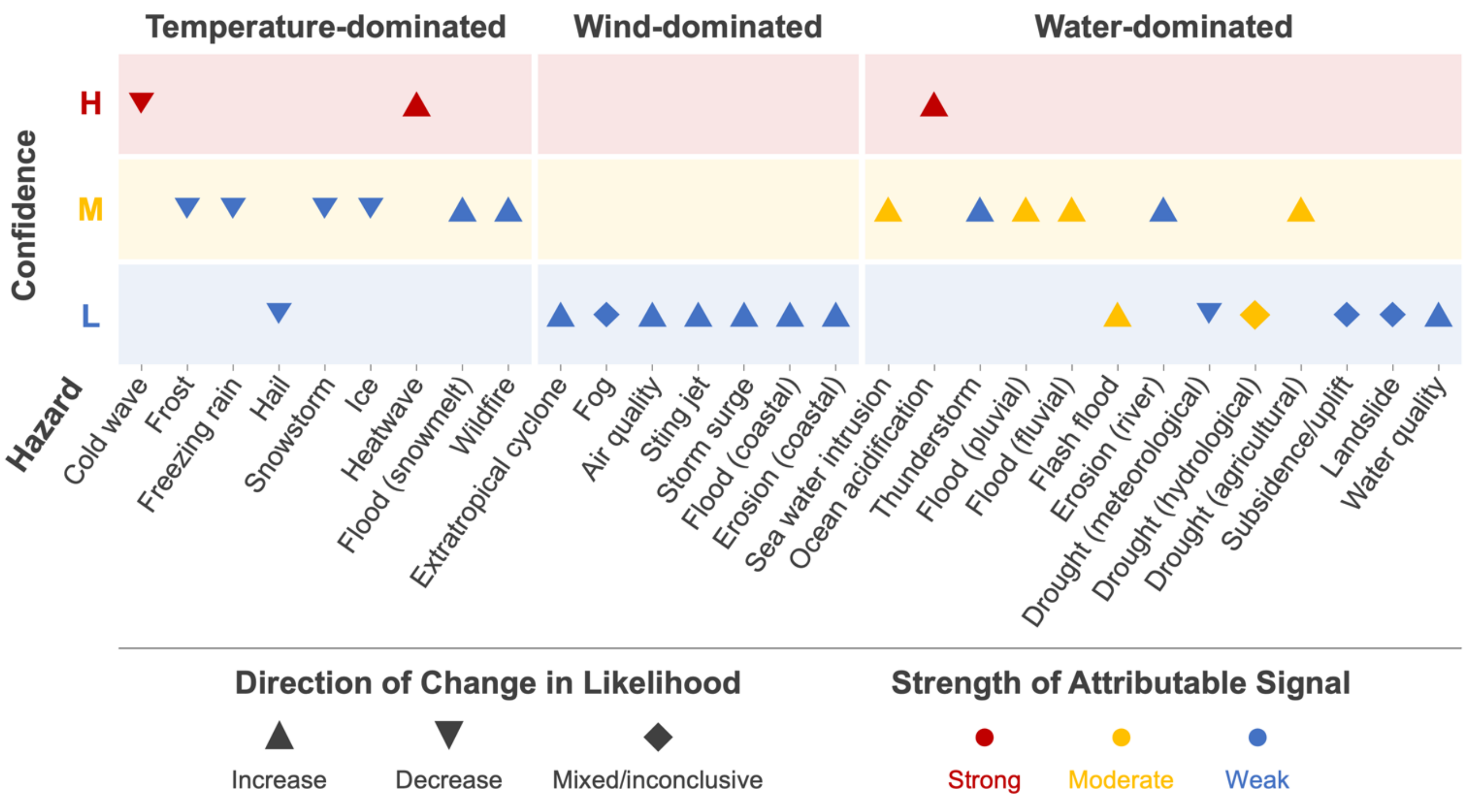

*Figure 1 – For each of the 29 hazards assessed in this review, this is a synthesis of the direction of change, strength of an anthropogenic climate change signal (attribution), and our confidence in that attribution, represented respectively by the arrows, their colour, and their placement in the rows. For example, for drought (hydrological), very few attribution studies relevant for this hazard were found (3). Two trend attributions for northern Europe, including the UK, found a moderate contribution of human influence to an increase in annual precipitation and river flows in the last century, though heavily modulated by the North Atlantic Oscillation. Whereas a study for a specific drought event found a strong human contribution to a notable rise in the likelihood of a hot, dry summer since the 1970s. Due to the low number of studies for this hazard and their differing conclusions, depending on the attribution method, our confidence in any concluding attribution statement was low. Further details are in Chapter 6.*

# Table of Contents

# 1 Background to the Review

The UK's climate is changing. Our uniquely comprehensive long-term observational record displays an upward trend in temperatures since the 20th century (Karoly and Stott, 2006; Slingo, 2021), both on average and within seasons and regions. Trends in average precipitation are harder to identify (McCarthy et al., 2021), especially regionally. Nevertheless, the magnitude and frequency of “extreme” precipitation and temperature events are also changing (Christidis et al., 2010). As an intrinsically complicated and chaotic system, our climate experiences natural variations. However, anthropogenic (human) activities are now accepted to be the main driver behind recent changes in large-scale temperature and precipitation over multi-decadal timescales (IPCC, 2021). Changed climate and weather can in turn manifest as environmental hazards which impact both natural and human systems.

In order to identify a change in an observation (detect) and understand the cause of that change (attribute), advanced analysis tools are needed. Detection and Attribution (D&A) studies seek to identify a trend, pattern, or change in climate-related variables or systems that can then be linked to anthropogenic drivers. Basic physical theory can indicate what changes we might expect. For example, in general, higher greenhouse gas (GHG) concentrations trap heat in the lower atmosphere, so should warm the surface. A warmer atmosphere should also hold more moisture, thus increasing the intensity of precipitation extremes in general (Fischer and Knutti, 2016; Trenberth, 2011). But whether that is the case everywhere depends on several factors, including natural internal variability (noise), natural forced variability (e.g., volcanic emissions, solar output) and the role of other anthropogenic forcings beside GHG emissions, such as land use changes. A change can only be attributed once all factors are considered and human influence remains the only reasonable mechanism driving that change. Relatively few D&A studies have attributed impacts on natural and human systems to anthropogenic climate change, as this is an emerging area of research. Impact attribution can be challenging as the full effect on systems depends not only on the climate-related hazard, but the system’s exposure and vulnerability too (see Section 1.4). The Intergovernmental Panel on Climate Change (IPCC) in 2022 published their report assessing the impacts of climate change at global and regional levels (IPCC Working Group II, 2022). They concluded that “[w]idespread, pervasive impacts to ecosystems, people, settlements, and infrastructure have resulted from observed increases in the frequency and intensity of climate and weather extremes” (Pörtner et al., 2022). Although some of the IPCC’s assessment included formal attribution of impacts to anthropogenic changes, they mostly attributed to climate change in a general sense. This contrasts the approach taken here, discussed further in Section 2.1.

Additionally, the Technical Report for the 3rd UK Climate Change Risk Assessment (CCRA3: (Betts et al., 2021)) did not go into detail on the topic of D&A, despite a key step in that report’s method being an assessment of the magnitude of current risks (Watkiss and Betts, 2021). This is discussed further in Chapter 2. This review is a synthesis of D&A studies covering the UK. It draws together evidence on climate change-driven trends and extremes within the observational record, in order to support and inform the next Climate Change Risk Assessment (CCRA4).

## 1.1 The Definition of Detection & Attribution

Detection is the identification of a signal or change, and attribution is the identification of its driver. However, there are a variety of definitions of, and approaches to, D&A, with some communities using the term very broadly, and others only referring to it when specific statistical methods are used.

Detection demonstrates that the climate, or system affected by it, has changed in some defined, statistical sense, without providing a reason (Hegerl et al., 2010; Stone et al., 2013). The IPCC stated that human-driven climate change is now “unequivocal” (IPCC, 2021) – or, in other words, it has been detected. To ensure detection, the signal (climate change) must be separated from noise (natural variability) at some statistical significance level, which requires an estimate of the system’s noise. For this, an observational record can be used if it is sufficiently comprehensive (Hegerl et al., 2010), or a long-running simulation of preindustrial-representative conditions without any evolving, historical drivers. Since we have now detected a change in many of our climate-related indicators, it may seem reasonable to assume that some of the individual weather events that comprise “climate” must also have changed. Indeed, over the last few decades, an increasing number of events have been attributed to climate change (Lloyd and Oreskes, 2018).

Attribution is used to identify the cause of the detected change. Attribution considers whether a specified set of external drivers, or forcings, are the reason behind that change (Hegerl et al., 2010). Typically, attribution requires a combination of statistical analysis and physical understanding to determine cause and effect. This may require many different sources of evidence, depending on the system (Hansen et al., 2016). The IPCC definition of attribution has evolved with their assessment reports (ARs); from a generic concept to a specific description of approach (Stone et al., 2013). However, even within the latest AR6, the different working groups (WGs) differ in that approach. In the cross-WG box on attribution, it is defined as “the process of evaluating the contribution of one or more causal factors to ... observed changes or events [in the climate system]” (Hope et al., 2022). WGI focuses on physical changes in the climate system so considers measurable, typically meteorological, indicators of climate change for attribution, such as air and ocean temperatures, precipitation, sea level, etc. (Eyring et al., 2021). Meanwhile WGII focuses on the impacts of climate change, with impacts on ecosystems and human communities mostly attributed to general climate change, not necessarily anthropogenic (O’Neill et al., 2022). In general, D&A studies fall into one of three main types (Zhai et al., 2018):

1. Trend Attribution: long-term spatial and temporal trends and patterns in climate variables,
2. Event Attribution: extreme weather and climate events (hazards),
3. Impact Attribution: climate-related consequences for natural and human systems (impacts).

## 1.2 Different Types in Attribution

The focus of attribution studies started with long-term meteorological changes in the 1990s, with global mean surface temperature being a very early example. This extended to specific extreme events in the 2000s (Stott et al., 2004), and impacts in the 2010s (Pall et al., 2011). Simultaneously, attribution recently began to be considered as a climate service and has developed to a point where analysis can be performed within days or weeks rather than months (Stone and Hansen, 2016; Stott et al., 2016). Despite the prevalence of studies, it remains difficult to achieve complete coverage spatially and of event/impact types, even with the move from global to regional scales (Stott and Christidis, 2023; Zhai et al., 2018).

Trend analyses evaluate the extent to which “fingerprints” of response to some external forcing in climate models can explain patterns in observations (Hegerl and Zwiers, 2011). Event attributions conversely aim to identify the extent to which anthropogenic climate change has changed an event’s frequency or magnitude. Impact attribution utilises some of the same techniques, though is less common, for reasons discussed below and in Sections 1.4 and 2.3.

### 1.2.1 Trend Attribution

Trend attribution aims to distinguish internally driven changes from those that are externally driven. They compare expected space-time patterns of response to some external forcing with “fingerprints” of response in observations (Hegerl and Zwiers, 2011). The earliest statistical method for this was simply pattern correlation but now tends to be linear regression (Allen and Tett, 1999; Hasselmann, 1993). This method assumes that the response to external forcing is a deterministic change, and that signals and noise superimpose linearly. This has been found to be valid for large-scale temperature change, but not other variables like precipitation, where trends are not always detectable and can be region dependent. Additionally, this approach is not necessarily appropriate for identifying the exact anthropogenic driver, nor for differentiating changes due to other, slowly varying external drivers (Hegerl and Zwiers, 2011).

### 1.2.2 Event Attribution

The concept of attributing a single extreme event to climate change was first formalised in Allen (2003) in response to flooding in Oxford in January 2003 (Allen, 2003). This was followed by the first event attribution analysis, using the European heatwave of 2003 as a case study (Stott et al., 2004). Event attribution studies estimate "the extent to which human-induced climate change has altered the likelihood of a particular threshold relevant to the event being exceeded ... or the extent to which climate change has altered the intensity of an event." These represent two approaches. The "risk-based" approach that usually considers classes of events, and the "storyline" approach which focuses on a specific event.

The storyline approach aims to determine whether climate change caused an extreme event, so focuses on changes in magnitude (e.g., maximum temperature in a heatwave) (Lloyd and Oreskes, 2018; Stott and Christidis, 2023). The risk-based approach alternatively tends to calculate changes in likelihood of classes of events, defined by a threshold in a climate variable (e.g., rainfall), or in an impact (e.g., flood defence exceedance). It recognises that extreme events can have multiple external drivers (natural or anthropogenic) on top of the system's internal noise and that human activities can influence the likelihood of such events occurring. Such studies assess extreme events as they might have been in a world without human influence. With the relatively short observational record, they rely on climate model simulations with actual and "counterfactual" climate conditions, where a counterfactual world is a natural one without anthropogenic forcings, described in more detail in Section 2.3.1. This approach assumes that models can reliably simulate the event (Otto, 2017), discussed further in Section 0. Ultimately, these studies determine whether human influence made the event more or less likely, had no effect, or is inconclusive with current techniques.

### 1.2.3 Impact Attribution

Impact attribution utilises the techniques of both trend and, most commonly, event attribution. Non-impact studies often include contextual details about impacts, demonstrating that an attributable event can impact on natural and/or human systems, without specifically attributing it. Explicitly identifying climate change as driving an impact requires consideration of contributions from other, confounding factors, as in trend and event attribution. If there is no other reasonable mechanism besides climate change, then it has been attributed. However, for many impacts, non-climate drivers are equally, if not more, important than climate change (Mitchell et al., 2022). Accounting for these may mean that a notable change in a climate variable, such as rainfall, does not necessarily produce a notable impact, such as a destructive flood. Furthermore, attributing a change in climate variables and impacts both require a significant number of high quality observations over a sufficiently long period and at the appropriate resolution. For the UK's climate, that kind of data typically exists, but impacts can be on ecological and environmental or social and economic systems, so utilise many more lines of evidence. For these, there may be insufficient data, or it may require synthesis of interdisciplinary information that differs in terminology and type (Stone et al., 2013; Uhe et al., 2021a). Consequently, the full impact-to-anthropogenic climate change attribution chain cannot often be modelled easily. As a result, there are few impact studies, especially for the UK.

In "single-step" attribution, one modelling set-up is used to attribute a detected change (Hegerl et al., 2010). Whereas a "multi-step" approach links separate, single-step approaches into one overall attribution assessment (Stone et al., 2013). First, an observed change in a hazard or impact is attributed to changes in one or multiple meteorological variables. Then, that change is separately attributed to an external forcing, such as anthropogenic climate change (Zhai et al., 2018). When attributing rare events, or the impacts of them, it may not necessarily be possible to estimate whether its frequency or magnitude has changed from observations alone. However, it may be possible to perform a multi-step attribution assessment of an indirectly estimated change in the likelihood of the event/impact (Uhe et al., 2021a). For example, the change in heatwave-driven excess deaths may not itself be attributable, but attribution of extreme temperature can infer the role of anthropogenic climate change in the hazard and/or impact (Ebi et al., 2017; Hegerl et al., 2010). All impact D&A studies included in this review use the multi-step approach.

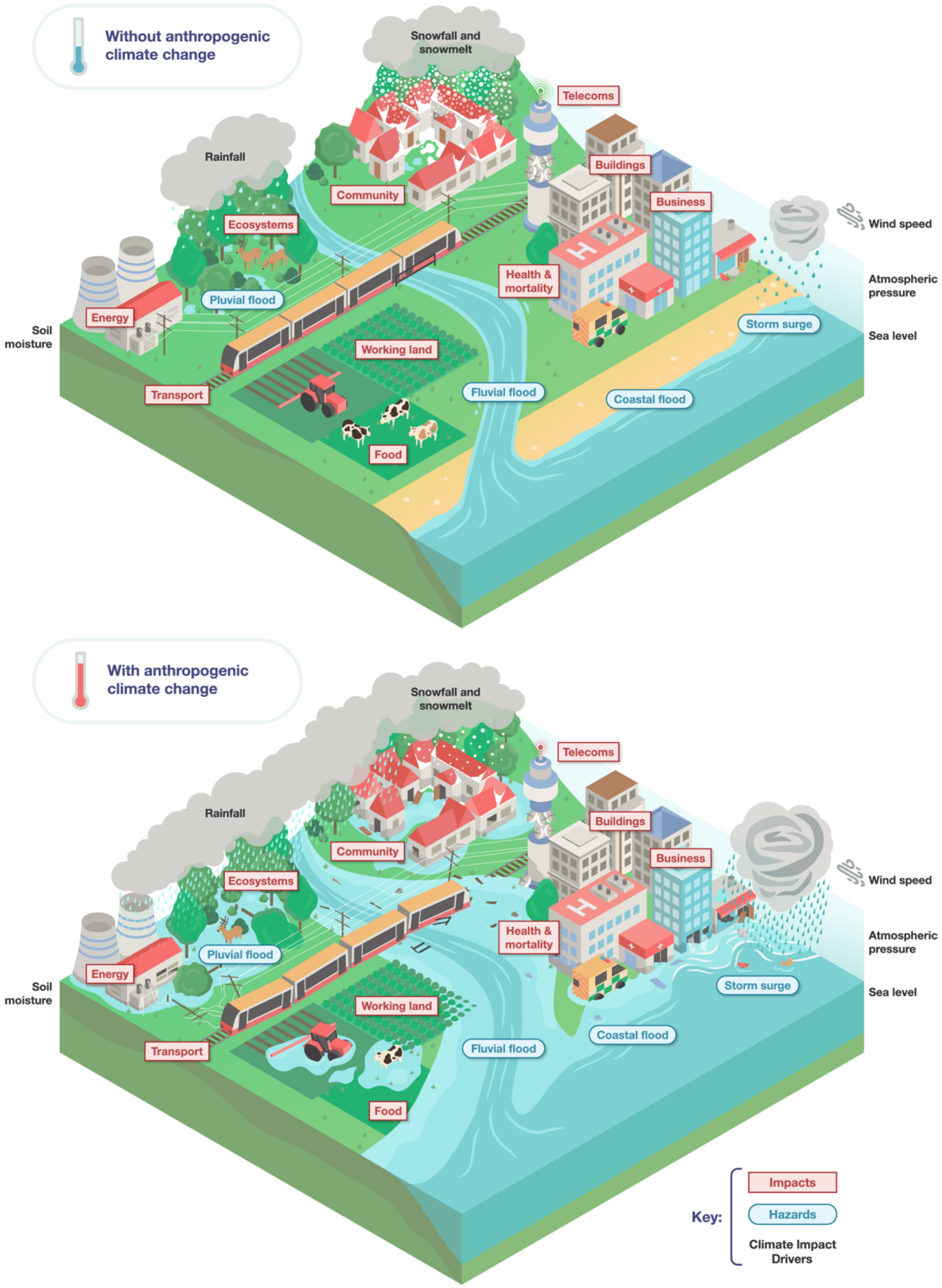

*Figure 2 – An attribution example for flood-related natural hazards (blue text) via associated CIDs (black), with some of the key impacts on human systems (red). Natural weather and climate variability can lead to natural hazards and impacts even under pre-industrial conditions (e.g., upper panel), including severe events and impacts, though these may be relatively rare. Anthropogenic climate change alters many meteorological variables (CIDs) that may exacerbate such hazards and impacts (e.g., lower panel), increasing their frequency, intensity, and/or extent. The key task of attribution is to distinguish between the two in order to robustly quantify those changes in hazards and their impacts that are due to climate change.*

## 1.3 The Difference between Hazards, Impacts & Climate Impact Drivers

There is currently no common definition to delineate hazards, impacts, and Climate Impact Drivers (CIDs). We choose to define them as follows, with Figure 1 demonstrating the interconnection between them. A **CID** is a physical condition of the climate system, and as such is a measurable meteorological variable. A (natural) **hazard** refers to how one or multiple CIDs affect some part of the Earth system. Different features of CIDs (e.g., sign, magnitude, temporal and spatial extent, frequency) can be important for different hazards. There can also be an interdependence between hazards, such as between a storm surge, coastal flooding, and coastal erosion. Finally, an **impact** refers to the effect of hazards on natural and/or human systems. Most D&A literature is for CIDs and hazards, with relatively few impact attribution studies, as previously mentioned.

## 1.4 The Potential Confusion with Definitions of Risk

In climate science, "risk" is commonly viewed as a combination of hazards, exposure, and vulnerability of natural and human systems, as demonstrated in Figure 3. This is the definition used in CCRA3, as explained in the Technical Report's Method chapter (Watkiss and Betts, 2021). While socioeconomic factors are important for human systems, they are out of the scope of this review, with the focus instead on changes in the climate system as measured by CIDs, which are simpler to attribute.

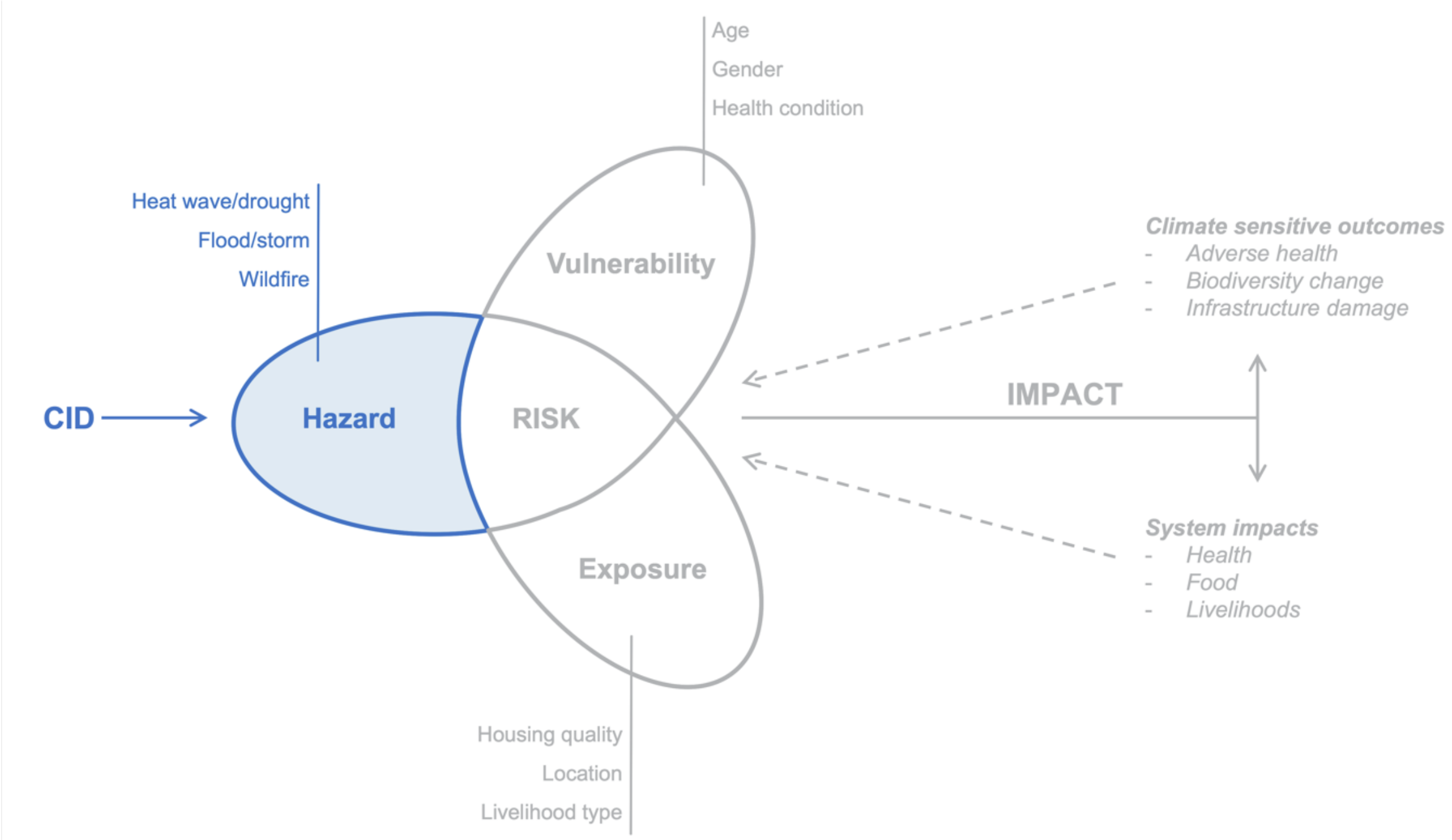

*Figure 3 – Conceptual diagram of risk as a function of hazard, exposure, and vulnerability. Blue highlights the areas of attribution in focus in this review; CIDs and associated climate hazards. Recreated from Fig. 7.4 in Cissé et al., 2022: "Interactions between hazard, exposure and vulnerability that generate impacts on health systems and outcomes, with selected examples".*

Alternatively in the D&A community, "risk" is an assessment of how likely an event is to happen. If the risk of an event is greater in a world with human emissions compared to one without, humans made that event more likely to happen. It is this definition of risk that will be used in this review. For example, the risk-based approach (see Section 1.2.2) assesses the likelihood of a certain type of event occurring due to climate change (Lloyd and Oreskes, 2018). It compares event probabilities from climate models with actual climate conditions ($P_{ACT}$), including natural and anthropogenic forcings, to those with counterfactual conditions to those with counterfactual conditions ($P_{NAT}$), including only natural forcings (Bellprat and Doblas-Reyes, 2016; Lloyd and Oreskes, 2018). Then, the risk ratio ($\mathrm{RR}$) and fraction of attributable risk ($\mathrm{FAR}$) indicate the extent of human influence on the change in an event's probability or intensity (Stott et al., 2016; Zhai et al., 2018):

$$\mathrm{RR} = \frac{P_{ACT}}{P_{NAT}}$$

$$\mathrm{FAR} = 1 - \left(\frac{1}{\mathrm{RR}}\right)$$

$\mathrm{FAR}$ is a probabilistic measure, with $\mathrm{FAR}$ > 0 indicating that an event has become more likely due to climate change (Bellprat and Doblas-Reyes, 2016). $\mathrm{FAR}$ can theoretically be infinity, and as such tends to be harder to interpret. Increasing degrees of freedom appear to reduce the $\mathrm{FAR}$ (Christiansen, 2015). For example, $\mathrm{FAR}$ can depend on the diagnostic chosen, number of ensemble members, and the size of the region and period studied. In addition, unreliable models have been noted to overestimate $\mathrm{FAR}$ due to overconfident ensemble spread, necessitating ensemble calibration and bias correction, which are themselves uncertain (Bellprat and Doblas-Reyes, 2016). Meanwhile, $\mathrm{RR}$ frames the result of an attribution investigation in terms of relative probabilities, so is applicable for both increases and decreases in event frequency (Wilcox et al., 2018). The greater the $\mathrm{RR}$, the greater the role of climate change; e.g., $\mathrm{RR}$ = 2 means that an event is twice as likely due to climate change, while $\mathrm{RR}$ = ∞ (i.e., $P_{NAT} = 0$) means an event could not have occurred at all without it (Vogel et al., 2019). In this review, $\mathrm{RR}$ is preferred due to its relative ease of interpretability. However, it should be noted that it depends on event definition, including domain (i.e., local vs. regional) (Uhe et al., 2016) and timescale (e.g., 1-day vs. season-long "events") (Leach et al., 2020). Furthermore, a change in the likelihood of an event may appear large but only correspond to a marginal change in magnitude, such that the real-life impact is relatively small. Many studies investigate the change in one or the other depending on the attribution question being asked (see Section 2.3.1), and as a result may not show the full picture. This can be circumvented by using classes of events, defined by a fixed magnitude threshold (Perkins-Kirkpatrick et al., 2022).

# 2 Approach to the Review

There was minimal discussion of D&A in the CCRA3 Technical Report (Betts et al., 2021). The climate science chapter referenced a few studies and there was some discussion of the approaches to and challenges in extreme event attribution (Slingo, 2021). However, while there were statements linking observed impacts to observed changes in climate in the chapters assessing sector-by-sector risks (Berry and Brown, 2021; Challinor and Benton, 2021; Jaroszweski et al., 2021; Kovats and Brisley, 2021; Surminski et al., 2021), there were few risks for which the evidence included studies of the specific role of anthropogenic climate change. Attribution can support public and private sector-led adaption to climate change. Comprehensive information on climate extremes and their impacts can aid policy and decision-making (James et al., 2014, 2019); combining with exposure and vulnerability enhances our understanding of "risks" to society and the natural environment. Attribution can also serve as the evidentiary basis for litigation concerning the impacts of climate change. An increasing number of lawsuits in recent years have aimed to compel governments and corporations to act against climate change (Stuart-Smith et al., 2021). Several lawsuits have aimed to hold certain organisations responsible for climate change impacts resulting from their GHG emissions (Kaminski, 2022). This requires a demonstration to the courts of a causal relationship between the defendants' emissions and the plaintiffs' losses due to climate-related events – this can be provided by an attribution assessment.

This review of D&A acknowledges the need for a better understanding of the changes in climate, weather, and hazards across the UK ahead of CCRA4. Detection highlights emerging trends and attribution explains why and how much hazards have changed, which can aid mitigation and adaptation. Focus on recent events and present-day likelihood of extremes enables us to assess current vulnerabilities to inform actions in the near term. The rest of this chapter will detail our approach to the review, including our criteria for inclusion, and some of the limitations worth bearing in mind.

## 2.1 What's in Scope: D&A Studies

D&A studies included in this review must meet the criteria summarised in Table 1. As discussed in Section 1.4, hazards are just one contributor to impacts on natural and human systems. While socioeconomic changes are important for human systems, they are out of the scope of this review, with the focus instead on changes in the climate system as measured by CIDs. Furthermore, studies that fully attribute a hazard or impact to anthropogenic climate change are preferred, but not required.

| Category | Criterion | Required? | Details |
|---|---|---|---|
| **Detection** | High signal-to-noise | Y | i.e., signal-to-noise greater than 1 |
| | Low signal-to-noise | N | i.e., signal-to-noise less than 1 |
| **Attribution of …** | CID | Y | |
| | Hazard | N | Not required but useful if available |
| | Impact | N | |
| **Component of Risk** | Hazard | Y | |
| | Exposure | N | Socioeconomic factors are out of scope |
| | Vulnerability | N | |
| **Relevance to UK** | Direct | Y | i.e., the UK and its four nations |
| | Indirect | N | Climate change elsewhere is out of scope |
| **Record** | Past | Y | Recent past, i.e., direct observations |
| | Current | Y | |
| | Future | N | Future projections are out of scope |

*Table 1 – Criteria for inclusion of a study in the review: Y – required, N – not required and/or out of scope.*

The trend or event must be within the current or recent past observational record. This requires direct rather than proxy observations, such as the daily Central England Temperature (CET) record from 1722 onwards. Similarly, the study must demonstrate separation of signal (climate change) from noise (natural variability) – a study that fits a trend line is not sufficient. This may mean no detection for some variables but does not mean no change at all, as there could be a detectable impact in the near future. Whilst we also acknowledge the need to provide an assessment of attributable changes in CIDs with spatial disaggregation across the UK, and as such require a minimum level of England, Northern Ireland, Scotland and Wales, note that this has not always been possible due to a lack of studies, for Northern Ireland particularly. This may ultimately inform mitigation and adaptation strategies, but the power of the devolved administrations is likely to be limited by the data available.  This also means that climate change occurring in or affecting other countries is out of scope, although we recognise that the UK can be affected by such indirect change. Finally, we will consider studies that aim to attribute to anthropogenic climate change, though some may ultimately be unable to identify it as a dominant driver.

## 2.2 What's in Scope: Impacts, Hazards & CIDs

The key systems that can be impacted by climate change are listed in Table 2 and were informed by sectors from the upcoming Climate Change Committee Adaptation Progress Report (Dooks, 2023) and Table 12.2 in the WGI contribution to IPCC AR6 (Ranasinghe et al., 2021). In scope hazards are in Table 3, and were selected due to their recorded incidence or impacts. For some, this is based on the prevalence of attribution studies, though for most relates to the perceived, as well as actual, impacts in the UK. They are categorised as temperature, wind and water-dominated, within which they are listed in terms of increasing complexity. Finally, the CIDs in scope are in Table 4. We focus on these, as they are measurable (from observations and models) and were deemed relevant to the hazards.

| Impact System | Details |
| --- | --- |
| Ecosystems | Terrestrial, freshwater, ocean, coastal |
| Working Land and Seas | Agriculture, fisheries, forestry, soil |
| Food | Quantity, quality, accessibility, infrastructure |
| Water | Quality, accessibility, infrastructure |
| Energy | Infrastructure, availability |
| Transport | Infrastructure, availability |
| Telecoms and ICT | Infrastructure, availability |
| Built Environment | Buildings, towns and cities, infrastructure |
| Heritage | Buildings, people |
| Community | People, buildings |
| Health and Wellbeing | People, infrastructure, availability of services |
| Mortality | People, impact on services |
| Business | Buildings, people (e.g., labour productivity) |
| Finance | Markets, banking, investment, insurance |

*Table 2 - Impact systems in scope.*

| Driver | Hazard | Definition |
| --- | --- | --- |
| **Temperature** | Cold wave | A period of unusually low near-surface air temperatures |
| | Frost | A layer of ice generated by the deposition of water vapour |
| | Freezing rain | Rain droplets below $0℃$ |
| | Hail | A form of precipitation, consisting of ice particles |
| | Snowstorm | A heavy snowfall, often accompanied by high winds |
| | Ice | Frozen, solid water |
| | Heatwave | A period of unusually high temperatures compared to records |
| | Flood (snowmelt) | Results from the melting of snowpack built up over winter |
| | Wildfire | Any uncontrolled fire affecting a range of landscapes |
| **Wind** | Extratropical cyclone | A low-pressure system that develops outside of the tropics |
| | Fog | The suspension of $\mu$m-scale water droplets in the air |
| | Air Quality | The condition or level of pollution of air |
| | Sting jet | A core of high winds that can form in low pressure areas |
| | Storm surge | An unusually high water level |
| | Flood (coastal) | Results from a storm surge/high winds and high tides |
| | Erosion (coastal) | Removal of material from coastal land |
| **Water** | Sea water intrusion | The process by which saltwater infiltrates an aquifer |
| | Ocean acidification | The process by which the ocean is made more acidic |
| | Thunderstorm | A convective storm characterised by a sharp/rumbling sound |
| | Flood (pluvial) | Results from high rainfall |
| | Flood (fluvial) | Results from a rise in the water level of a body of water |
| | Flash flood | A highly localised flood, in time and space |
| | Erosion (river) | Removal of material from riverbanks |
| | Drought (meteorological) | A lack of precipitation |
| | Drought (hydrological) | A period of low water supply |
| | Drought (agricultural) | Drought that impacts agricultural production |
| | Subsidence/uplift | Lowering/raising of the ground |
| | Landslide | Movement of a mass of rock, debris, or earth downslope |
| | Water quality | The condition of water in relation to a specific use |

*Table 3 – Hazards in scope.*

| Type | CID |
|---|---|
| **Temperature** | Mean air temperature (BL) |
| | Extreme air temperature (BL) |
| | Air temperature (upper troposphere) |
| **Precipitation** | Mean rainfall |
| | Mean snowfall |
| | Extreme rainfall |
| | Extreme snowfall |
| | Evaporation |
| | Humidity (specific & relative) |
| **Wind** | Mean wind speed |
| | Extreme wind speed |
| **Land** | Soil moisture |
| **Marine** | Relative sea level |
| | Mean ocean temperature |
| | Extreme ocean temperature |
| **Other** | Lightning |
| | Solar radiation (BL) |
| | Atmospheric pressure (BL) |

*Table 4 – Climate Impact Drivers (CIDs) in scope, where BL refers to boundary layer (i.e., near-surface).*

## 2.3 Assumptions & Limitations

The studies covered in Chapters 3 to 5 were found predominantly through discussion with experts, previous literature reviews (e.g., IPCC AR5 and 6), web of science searches, and forwards/backward citations. It is therefore possible that not all relevant literature has been included. Furthermore, most hazards in scope have no UK relevant D&A studies. This is likely due to a lack of observational data, as well as the fact that most models cannot resolve relevant processes, including land surface processes that are relevant for hazards such as erosion, or the use of parameterised convection which limits analysis of storm-related hazards, such as storm surges or sting jets. In this review, similar to the IPCC, assessments of “confidence” are based on the evidence (type, quantity, quality, consistency) and agreement between them. In our case, the evidence is drawn from the D&A literature.

Additionally, D&A relies on a variety of assumptions, depending on the approach taken and question asked. The points raised in the following sections should be kept in mind throughout the review.

### 2.3.1 Dependency on the Question Asked

Attribution strongly depends on the framing of the initial question. For example, an observations-based approach asks: “How has the frequency and magnitude of a certain class of extreme events changed over time?” While a modelling approach asks: “What is the role of anthropogenic climate change in the frequency and magnitude of an extreme event occurring in the current climate?” (Otto, 2017). These use simulations of actual and counterfactual climates to compare the world as it is to what it might have been without human influence (Zhai et al., 2018). The anthropogenic climate change itself varies between studies; a counterfactual climate may be produced by removing a single external driver, removing multiple drivers, or proxy preindustrial conditions with an earlier observed or simulated period. The question also depends on the climate models themselves. Atmosphere-only and fully coupled atmosphere-ocean general circulation models (AGCMs and AOGCMs, respectively) are associated with “conditional” and “unconditional” attribution, respectively. The latter samples the range of possible states of the ocean, while the former prescribes sea surface temperature (SST) and sea ice concentration (SIC), which can help reduce model biases and are comparatively computationally cheap to run, but may poorly reproduce extreme events that are influenced by atmosphere-ocean interactions (Stott et al., 2016; Zhai et al., 2018). As AGCMs tend to simulate forced changes in atmospheric composition, the pattern of warming that must be removed from the present day-representative SST/SICs to achieve the counterfactual conditions can also be generated in different ways, with some atmosphere-only model studies using SST/SIC patterns generated from a few different coupled models to account for uncertainty in ocean warming (Vautard et al., 2016). Some studies take such “conditional” attribution further and compare the relative contributions of anthropogenic climate change and climatic conditions by recreating the weather regimes/systems or circulation patterns observed during an event (Stott et al., 2016). Such studies ask: “What is the role of anthropogenic climate change in the frequency and magnitude of an extreme event occurring in the current climate, given that the atmosphere is in a specific state?” This approach can be useful for more regional studies.

### 2.3.2 The Difficulty with Smaller Scales

Trends in local or regional climate are not necessarily due to anthropogenic climate change and can be due to other factors. They can contradict rather than enhance the global signal or not emerge at all. Thus, warming of the globe does not indicate warming on smaller scales (Lloyd and Oreskes, 2018), and nor does it necessitate trends in other (non-temperature) variables, even if some are implied. Regional studies can have large uncertainties in forcings relevant at smaller scales, such as aerosols and land use change (Hegerl and Zwiers, 2011), as well as in response. Individual and local climate records have much greater natural variability than aggregated or global measures. This predominantly relates to the effects of averaging of noise on different scales, though factors including topography and turbulence, water and land management, and atmospheric circulation patterns or multidecadal natural variability can also influence local trends more than global ones (Hansen et al., 2016). Extreme events in specific locations are rare, so systematic changes in frequency may be undetectable in the observational record (Stott et al., 2016). Model-based methods can help overcome this. Models must be able to well represent physical processes and mechanisms linked to such extremes, including relatively realistic frequency of associated circulation patterns (Mitchell et al., 2017), and to represent the modes of natural variability that tend to drive regional extremes (Stott et al., 2016). There can also be a lack of spatial and temporal detail in the signal to be able to delineate the different forcings behind the regional trends or events (Hegerl and Zwiers, 2011). Plus, attribution of events like droughts, storms, or floods require high resolution models, inclusion of relevant surface processes, and simulation of impacts (Stott and Christidis, 2023). For example, for floods, this may mean using climate models, run-off models, river flow models, and flood inundation models, to capture the full extent of climate-to-flood impact (Uhe et al., 2021a). Furthermore, hazards like drought or wildfire are affected by multiple CIDs (precipitation, temperature, wind speed, soil moisture). Surface feedbacks then become just as important to understand and simulate as, for example, a warming trend. Impact attributions are more commonly performed at small scales, driven by availability of data for the impact system.

### 2.3.3 Having Confidence in Attribution

Confidence in an attribution statement is subject to the quality of and biases in the data. A favoured approach is to use multiple, independent D&A methods; if they reach similar conclusions, this minimises their individual uncertainties and enhances our confidence in the attribution (Zhai et al., 2018). Empirical approaches can use observations to estimate how much climate change has altered the risk of certain events. However, observational data is often only sufficient for validating model performance or verifying simulated results (Stott et al., 2016). In the former, these "bias corrections" can be simple when distributions match but are just offset by some fixed amount. However, given the focus on extreme events, matching extremes can be more difficult and varies between studies. Defining events or classes of event using return time, rather than absolute magnitude, is an alternative that avoids having to apply a bias correction based on extremes (Otto, 2017). In addition to bias correction, uncertainties in climate models can be introduced in the model structure, the construction of counterfactual conditions (see Section 2.3.1), and radiative forcing (Lane et al., 2022; Stott and Christidis, 2023). Similarly, "method uncertainty" in modelling can for example have substantial impacts on estimated precipitation changes (Uhe et al., 2021b). GCM results also sometimes feed into more specific hydrological or regional climate models, as mentioned in Section 2.3.2. These tend to have greater resolution but have other limitations. For example, not all hydrological models consider snow processes, which has been found to be important for fluvial flooding in upland areas of the UK (Bellprat and Doblas-Reyes, 2016). Furthermore, different precipitation and warming responses in a GCM can produce different river flow changes in a hydrological model, which can itself be highly dependent on resolution (Kay et al., 2011; Lane et al., 2022).

## 2.4 Organisation of the Review

The next three chapters are the literature review and present the current state of D&A studies relevant to the UK. 67 studies are covered here; some looked at individual hazards or related CIDs and others at hazards across multiple categories (13), mostly due to the overlap between wind and water-dominated hazards such as cyclones, storms, and flooding (7), and temperature and water-dominated hazards related to flooding (3), or heat and drought (2). Three studies looked at multiple types of event in a single year, such as heatwaves and floods.

The hazards in Table 3, categorised into temperature, water, and wind-dominated, will be introduced one at a time in each of the respective chapters. For every hazard, we will first provide background information for the hazard; definition, relevance to the UK, key impacts (as per Table 2) and expected change with anthropogenic climate change. Then the key CIDs will be listed, before the literature is reviewed, with particular focus on the UK and its four nations. A summary of the literature and any gaps identified will be raised last. In Chapter 6, we will summarise our findings, including knowledge gaps and suggested data or methodological advances required to address them.

# 3 Temperature-dominated Hazards

## 3.1 Cold Wave

### 3.1.1 Background

A cold wave is a period of unusually low near-surface air temperatures during the cold season, characterised by a significant drop in temperature and persisting for at least two consecutive days (Murray et al., 2021). Most days in a year are considered “cold” in the UK, resulting in an estimate average of approximately $60{,}000$ cold-related deaths each year in England and Wales alone (Gasparrini et al., 2022). Epidemiological studies show a U or V-shaped relationship between temperature and mortality in the UK; the risk of mortality increases substantially as temperatures drop below $5-6$℃ (Charlton-Perez et al., 2019). Water, food, and energy systems can also be impacted by prolonged low temperatures, with a similarly U-shaped relationship for demand in the latter (Christidis et al., 2021b). The 21$^{st}$ century has so far been warmer than any other period of equivalent length within the last $300$ years, with 2012 – 2021 winters $1.2$℃ warmer than that of 1961 – 1990 (Kendon et al., 2022). Thus, with anthropogenic warming, the UK is expected to see milder winters on average – i.e., fewer cold waves.

### 3.1.2 Key CIDs

The key CIDs are extreme low near-surface air temperature and increasing mean and extreme high snowfall. Short-lived heavy snowfall and multi-day build-up of snow can have significant impacts on spatial scales from sub-km to $10$s of km. Whereas, cold temperature-related impacts on health and mortality have been reported to last for up to weeks at a time following an event and during extended cold waves (Gasparrini et al., 2015).

### 3.1.3 Literature

There are relatively few D&A studies, but all find that the severity of cold extremes is lessening.

#### *3.1.3.1 UK*

The cold spring of 2013 notably affected energy, agriculture, and businesses. Using AGCMs with and without anthropogenic forcings, Christidis et al., 2014 found that anthropogenic climate change had reduced the odds of such an event by at least $30$ times. This is complemented by a study of the unusually warm 2015/16 winter, which found that anthropogenic forcings increased the risk of exceeding warm winter temperatures ($> 20$℃) by at least a factor of $3-4$ (Christidis et al., 2018). Relatedly, an assessment of extreme UK winter sunshine using atmosphere-only models showed that human influence, particularly reduced aerosol concentrations in recent decades, increased the risk of extreme winter solar flux events by a factor of $3.6$, but noted that the record sunshine in winter 2014/15 occurred within a meteorological context atypical of sunny conditions (Christidis et al., 2016). Christidis et al., 2021b also assessed the impact of anthropogenic climate change on domestic energy consumption over 2008 – 2019 by comparing fully coupled models of current and counterfactual climate, though assuming the same level of adaptation. They found that without anthropogenic influence, UK households would approximately consume an extra $1{,}400$ kWh/year. In other words, anthropogenic warming lessened the need for people to heat their homes.

#### *3.1.3.2 England*

Massey et al., 2012 performed a trend attribution to investigate whether the odds of a warm November and cold December in England had changed. They used the CET record with AGCM simulations of two climate scenarios representative of the 1960s and 2000s, bookending the most significant period of human influence on climate. They found that warm temperatures in November became $62$ times more likely and a cold December ($< -0.7$℃) half as likely to occur in the later period. They noted that the temperature of a 100-year event in November had increased by $1.45$℃ between the 1960s and 2000s, and in December by $0.87$℃.

For winter 2010/11, Christidis and Stott, 2012 used long-term observations and AGCMs with and without anthropogenic forcings. They found that human influence had reduced the risk of such cold December and January temperatures by at least $20\%$. The same authors later studied the cold spring of 2018, which had notable mortality and health impacts. Using long-term observations and coupled models this time, they also found that such events had become less likely with human influence, with a $\mathrm{RR}$ of at least $5.84$ in favour of a world without human influence (Christidis and Stott, 2020). This was corroborated by their later study of the unusually warm 2018/19 winter. Such an event corresponded to a $+5.2℃$ anomaly in the present, compared to $+4.4℃$ in the early twentieth century, with $\mathrm{RR} = 282$ from a coupled model-based analysis of warmest winter day in central England (Christidis and Stott, 2021).

Finally, Charlton-Perez et al., 2019 investigated cold-related impacts on the health system during English winters in 1900 – 2016 using AGCMs and AOGCMs with and without anthropogenic forcings. The found that the risk of high demand days had been approximately halved by anthropogenic forcing, with $\mathrm{RR} = 0.34$ and $0.52$ when comparing the 2000 – 2016 climate to pre-industrial and 1960 – 1990 climates, respectively. However, they also noted substantial year-to-year variability in attributable mortality, as well as extreme low temperatures.

### 3.1.4 Summary & Gaps

The studies reviewed here indicate that human influence has reduced the frequency and extremity of cold extremes in winter and early spring, with a high attributable signal. These UK and England-focused assessments corroborate larger-scale temperature trend attribution studies, discussed further in Section 3.7, giving us high confidence in the attribution. Two of the above studies attribute a lessening of pressure on the health system and energy services in recent decades to climate change, but cold tends to dominate temperature-related impacts on a range of UK sectors and we are still missing a complete understanding of the changing impact of cold extremes. Ensuring accurate and reliable impact attribution is especially important as model biases in temperature extremes have been found to be larger than that of seasonal averages, with cold extremes sometimes simulated too cold (Wehrli et al., 2018).

## 3.2 Frost

### 3.2.1 Background

Frost is a layer of ice generated by the deposition of water vapour from surrounding air (Murray et al., 2021), which can affect agriculture, and thus food systems. As for cold waves, with mean temperatures in the UK expected to rise with climate change, frosts may become less frequent. 2012 – 2021 already saw $18\%$ fewer days of ground frost compared to 1961 – 1990 (Kendon et al., 2022).

### 3.2.2 Key CIDs

The key CIDs are extreme low near-surface air temperature, low evaporation, and high atmospheric humidity and soil moisture. These all generate impacts on daily and sub-km scales.

### 3.2.3 Literature

There are no D&A studies that attribute changes in characteristics, occurrence, nor direct or indirect impacts of frost in the UK.

### 3.2.4 Summary & Gaps

The expected change in the key CID of extreme low temperature (see Section 3.1) is in line with recorded changes in frost since the 1960s. Resultant changes in related impact systems, such as agriculture, could be more beneficial than detrimental – thus, the lack of D&A studies is not considered a gap.

## 3.3 Freezing Rain

### 3.3.1 Background

Freezing rain occurs when the temperature of rain droplets is $< 0℃$. It forms when liquid water falls through a layer of sufficiently thick and cold air which can cause the rain to be supercooled and freeze on impact with any object, forming a layer of ice (Murray et al., 2021). It can affect transport, energy, and telecoms and ICT infrastructure, for example through power transmission. However, with anthropogenic warming, there are likely to be fewer instances of cold extremes in the UK (see Section 3.1), and therefore freezing rain.

### 3.3.2 Key CIDs

Its associated extreme low near-surface air temperature, increasing mean rainfall, and extreme high rain and snowfall can cause notable impacts over daily and sub-km scales.

### 3.3.3 Literature

There are no D&A studies that attribute changes in freezing rain nor its impacts in the UK.

### 3.3.4 Summary & Gaps

As freezing rain is a relatively rare occurrence in the current climate and, given the trend in cold waves (see Section 3.1), this is not considered a gap.

## 3.4 Hail

### 3.4.1 Background

Hail is a form of precipitation consisting of spherical, conical, or irregular shaped ice particles (hailstones) falling from a convective cloud and exceeding a maximum diameter on any axis of 5 mm (Allen et al., 2019; AMS, 2012). They form via riming or accretion of supercooled liquid water onto embryos of, typically, graupel, frozen drops, or other particles moving with updrafts (Allen et al., 2019). Growth depends on the time spent in the cloud layer amidst sufficiently supercooled water (approximately $-10$ to $-30℃$). Hail can damage buildings, energy, transport, and ICT and telecoms infrastructure, with downstream impacts on finance via insurance claims. Hail formation is complex, so its expected change with anthropogenic warming is uncertain; moisture and convective instability is expected to rise, suggesting a higher chance of hail (Raupach et al., 2021), but the necessary cold upper cloud temperatures may become less common.

### 3.4.2 Key CIDs

The key CIDs are anomalous warm near-surface and cold upper troposphere temperatures, and extreme high wind speeds and precipitation, which can cause notable impacts on sub-daily and sub-km scales.

### 3.4.3 Literature

There are currently no D&A studies that attribute changes in characteristics, frequency, nor direct or indirect impacts of hail in the UK.

### 3.4.4 Summary & Gaps

Given the potential for damage, the lack of D&A studies is a knowledge gap. Climate models are unable to resolve the microphysical processes involved in producing hail without being highly computationally expensive (Brimelow et al., 2017). Nevertheless, there have been a limited number of non-attribution studies that document a shift in hailstones' size distributions in recent years (Dessens et al., 2015), while the number of damaging hailstorms in the UK are projected to reduce (Sanderson et al., 2015). These studies' methods could be leveraged for attribution of changes in hail events in the UK.

## 3.5 Snowstorm

### 3.5.1 Background

Snowstorms are associated with heavy snowfall and, often, high winds (Murray et al., 2021). They can impact human health and communities, as well as water, energy, transport, and ICT and telecoms systems. With anthropogenic warming, the UK is expected to see milder winters (see Section 3.1) and a reduction in snowstorms would be a continuation of the trend of fewer and less severe UK snow events since the 1960s (Kendon et al., 2022).

### 3.5.2 Key CIDs

The key CIDs are extreme low near-surface and upper troposphere temperatures, extreme high wind speed and snowfall, and increasing mean snowfall. Multi-day build-up of snow, as well as the sub-daily to daily impact of extreme temperatures, winds, and heavy snowfall, can cause significant impacts over $10$s of km.

### 3.5.3 Literature

As for hail, there are no D&A studies specifically for snowstorms in the UK. Here, we look at two studies for UK cold waves which noted snowfall during the event. Attribution of a related CID such as cold temperatures can provide insight into the human influence on snowstorms.

#### *3.5.3.1 UK*

Spring 2013 saw extremely low temperatures accompanied by heavy, unseasonal snowstorms. In some areas, the snowfall led to livestock and financial losses and was associated with damage to energy infrastructure in Scotland and Northern Ireland. Using atmosphere-only models with and without anthropogenic forcings, Christidis et al., 2014 found that anthropogenic climate change reduced the risk of such a cold event by at least $30$ times.

#### *3.5.3.2 England*

In spring 2018, snowstorms resulted from the combination of a cold Siberian weather system, with a sudden stratospheric warming (SSW) driven Arctic air mass, and a winter storm over the UK. Another event attribution using long-term observations and fully coupled models found that the extreme low temperatures were also not attributable to anthropogenic climate change. Instead, $\mathrm{RR}$ for the likelihood of cold extremes *without* human influence was found to be at least $5.84$ (Christidis and Stott, 2020).

### 3.5.4 Summary & Gaps

Though there are no D&A studies specifically for snowstorms in the UK. The two studies above (and more discussed in Section 3.1) indicate that UK cold extremes have become less likely with anthropogenic warming, implying a similar reduction in conditions conducive to snow. Given that snowstorms in the UK can be highly disruptive, lack of D&A studies is a knowledge gap.

## 3.6 Ice

### 3.6.1 Background

Ice is frozen, solid water, and “icing” is a deposit or coating of ice on an object (Murray et al., 2021). Ice can impact water, transport, energy, and telecoms and ICT infrastructure, as well as human safety. The number of icing days, in which the daily maximum temperature stays $< 0℃$, has been decreasing since the 1960s in line with anthropogenic warming (Met Office, 2019a) – a trend that is expected to continue.

### 3.6.2 Key CIDs

The key CIDs are extreme low near-surface air temperature and sunshine, increasing mean rain and snowfall, as well as extreme high rain and snowfall. These can have impacts over sub-daily to multi-day timescales, and at sub-km to $10$s of km.

### 3.6.3 Literature
There are no formal attribution studies for ice in the UK.

### 3.6.4 Summary & Gaps
As in previous sections, ice constitutes a cold temperature-related hazard. Therefore, changes in the occurrence and magnitude of cold extremes as described in Section 3.1 can indicate the direction of change in ice-related hazards and impacts too (Massey et al., 2012). But icing can have wide reaching consequences, and there are currently no specific D&A studies for this hazard. Thus, how ice-related impacts have changed with anthropogenic warming is a gap.

## 3.7 Heatwave
### 3.7.1 Background
A heatwave is a period of unusually high temperatures, compared to records specific for the time of year and/or region, persisting for at least two consecutive days (Murray et al., 2021). Their impacts are wide-ranging, and most notably are on health, wellbeing, and mortality, as well as water and food systems. On average, approximately $800$ excess deaths are attributable to heat in England and Wales every year (Gasparrini et al., 2022), and heat impacts on health are more acute (within a few days) than that of cold. In the UK, heatwaves are becoming more common and severe, and are the topic of most UK D&A studies. This is partly due to the relative ease with which increasing average and extreme temperatures can be attributed to human-driven global warming. As the world continues to warm, “extreme” heatwaves are projected to become a fixture of UK summers (Lo and Mitchell, 2021), as well as features of other seasons, e.g., milder winters (Bloomfield and Thompson, 2023).

### 3.7.2 Key CIDs
The effects of high extreme near-surface temperatures can be exacerbated by both persistent high and low humidity, on sub-km and sub-daily to multi-day scales; heat-related health and mortality impacts are often immediate and can worsen within a few days (Gasparrini et al., 2015).

### 3.7.3 Literature
In the reviewed literature, the most common events assessed were the summer heatwaves of 2014 and 2018, with some winter heatwaves also included.

#### *3.7.3.1 UK*
There have been trend attributions for continental/hemispheric regions that cover the UK, and which provide a general picture of attributable changes in temperature. Using long-term observations and coupled models, historically rare hot summers in the northern hemisphere were found to become more frequent with human influence during the period 1900 – 2009, particularly in Europe (Jones et al., 2008). Various studies looked at trends over 1950 – 2009 using long-term observations, optimal fingerprinting methods, and coupled models with and without anthropogenic forcings. They found that human influence dominated global trends in mean surface temperature and the warmest night of the year, with northern Europe $\mathrm{RR} = 4.8$ and $2.7$, respectively (Christidis et al., 2010). One also noted an anthropogenic contribution to the increasing area affected by maximum and minimum temperature extremes in all regions, particularly from GHG forcing (Dittus et al., 2016). In addition, the risk of a year during 2000 – 2009 exceeding the warmest year since 1900 was estimated to have increased by at least a factor of $6$ in northern Europe (Christidis et al., 2012).

There are multiple event attribution studies for European heatwaves that affected the UK.
For example, Dong et al., 2014 investigated the hot, dry western European summer of 2013 using AGCM simulations with and without anthropogenic forcings. They reported that SSTs and sea ice explained $63\%$ of the area-averaged warming signal over western Europe, with the remaining $37\%$ due to changes in radiative forcings from anthropogenic GHGs and aerosols, which caused a substantial anticyclonic anomaly over the UK. The following year (2014) was then one of the warmest in Europe. Such a temperature anomaly was found to be at least $80$ times more likely for Europe in AGCMs with climate change, compared to those without (World Weather Attribution, 2014), with a strong anthropogenic contribution to long-term warming trends there too (King et al., 2015). Higher temperature variability was also noted on smaller scales; in the UK, the probability of such an event occurring was found to have approximately increased by a factor of $14$ (World Weather Attribution, 2014). Another study used four event attribution methods to study this event, including an empirical approach with historical data, coupled model ensembles with all and natural-only historical forcings, and very large ensembles of simulations from an atmosphere-only model. They reported that annual mean European temperatures experienced in 2014 were made at least $10 - 100$ times more likely by anthropogenic climate change. For regions on smaller scales and gridboxes in the centre of them, those lower bounds reduced to at least a factor of $4$ (Uhe et al., 2016).

June 2017 saw extreme high temperatures over western Europe. In a "real time" analysis, Otto et al., 2017 used AGCMs and AOGCMs with and without human GHG and aerosol emissions to study the event. In all studied countries they found that the risk of hot Junes had increased considerably, with past historical increases in GHGs raising the odds of a hot June in central England significantly ($\mathrm{RR} = 7$, or a change in magnitude of at least $0.7℃$). However, they also found a range of $\mathrm{RR}$s derived from observations and model simulations ($\mathrm{RR} = 4$ and $2.8 - 5.7$, respectively); as they ensured similar variability and corrected for biases in the mean, they reported that this was mainly due to differing estimates of the effect of anthropogenic emissions on summer temperatures in Europe.

In summer 2018, a high pressure anomaly across the north Atlantic/Europe associated with a north-shifted jet stream reduced the ability of cyclones to bring cool, moist Atlantic air to the UK and led to heatwaves across Europe (McCarthy et al., 2019). Vogel et al., 2019 investigated heatwaves occurring across the northern hemisphere that year using coupled model simulations. They found that a 2018-like event, with an average $22\%$ "concurrent hot day area" (covering the UK) between May and July, was not uncommon in present-day climate, but had little to no chance of occurring in model simulations of preindustrial conditions ($\mathrm{RR} \to \infty$) nor a period around $50$ years ago. Leach et al., 2020 focused on Europe and, using ACGM simulations of an actual and counterfactual world, found that for the UK, such season-long heat events were approximately 1-in-10-year events, with corresponding best-estimate $\mathrm{RR}$s between $10$ and $100$ for all regions. However, they noted that $\mathrm{RR}$s for the UK were likely to be overestimated as the prescribed SSTs underrepresented the intra-seasonal $2$-m temperature variability at almost all frequencies. McCarthy et al., 2019 alternatively used coupled models with observations for an event attribution for the UK specifically. They found that the present-day likelihood of a summer temperature anomaly at or above that in 2018 was approximately $30$ times higher than in a climate without anthropogenic GHG emissions. Kay et al., 2020 similarly used large ensemble coupled model simulations and found that the risk of exceeding summer 2018 temperatures was around $11\%$ in the current climate, which had risen rapidly from $< 1\%$ in the 1960s. These complement a study that used coupled models to examine the change in likelihood of extreme high UK temperatures ($> 30 - 40℃$). Christidis et al., 2020 found that a world affected by anthropogenic climate change had more areas likely to see temperatures exceeding $30 - 35℃$, with the return time for the $40℃$ threshold reduced from $100 - 1000$s of years in the natural climate to $100 - 300$ years in the present. Human influence was also found to increase the risk of setting an annual mean temperature record; using coupled models, the event in the 2018 climate was estimated to be approximately $120$ times more likely than in a world without anthropogenic forcings (Christidis et al., 2023).

The unusually high temperatures carried into winter 2018/19. Observations since 1960 indicated that this record-breaking winter heatwave was mostly due to the exceptional synoptic conditions present at the time, even without the influence of a warming climate. Nevertheless, they reported that anthropogenic climate change likely contributed to the heatwave's spatial extent and the margin by which the temperature record was broken (Kendon et al., 2020). Building on this, Christidis and Stott, 2021 used AGCMs and AOGMs with and without anthropogenic forcings and found that, while central England temperature extremes like in winter 2018/19 were rare even in today's warmer climate, they were still around $280$ times more likely with human influence. Leach et al., 2021 looked at the heatwave over Europe in February 2019 with a novel approach; they used an ensemble of forecasting models for a storyline attribution, with a variety of lead times to estimate the direct radiative impact of increased $CO_2$ concentrations (just one anthropogenic factor). They found that in their highly dynamically conditioned set-up, $CO_2$ directly increased the magnitude and the conditional probability of the heatwave in the UK – by a factor of at least $42$ for the latter. Then June and July 2019 saw extreme high temperatures over western Europe, including the UK, when certain weather patterns induced intense advection of hot air from north Africa. Vautard et al., 2020 used both AGCMs and AOGCMs for an event attribution study, though noted that changes in intensity were generally systematically underestimated compared to observations. Regardless, they reported that without human-induced climate change, summer heatwaves as exceptional as 2019 would be cooler; for the UK stations studied, the risk changed by at least a factor of $2 - 3$ and the magnitude of temperatures by around $1.5 - 2.5$℃.

An analysis with coupled models found that the risk of a year as hot as 2020 had increased by a factor of $50$ with human influence (McCarthy et al., 2021). Christidis and Stott, 2022a studied the extreme high temperatures in May 2020 for western Europe, including coverage of some parts of England, Wales, and Northern Ireland. They used coupled models with and without anthropogenic forcings to look at the influence of anthropogenic climate change and the anticyclonic circulation pattern present during the event. They found that anthropogenic forcings made May heatwaves at least as hot as 2020 around $40$ times more likely. They also noted that in the present climate, the persistent anticyclonic conditions, which tend to favour warm anomalies, made the event $2 - 4$ times more likely.

Finally, an investigation into global heat-related mortality using coupled models found that, at smaller scales, anthropogenic climate change attributable deaths were $< 1\%$ of the total warm season deaths for countries in northern subregions of Europe, including the UK (Vicedo-Cabrera et al., 2021). Nevertheless, across all global locations, heat-related mortality as a proportion of all warm-season deaths was an average of $0.58$ percentage points lower in counterfactual simulations.

#### *3.7.3.2 England*

One of the earliest D&A studies, by Karoly and Stott, 2006, used the CET record for 1700 – 2005 and coupled models with and without anthropogenic forcing for a trend attribution. They found that the almost $1$℃ warming in annual mean CET since 1950 was consistent with the response to the anthropogenic forcing. More recently, Hawkins et al., 2020 used long-term temperature records for Oxford, England for a fingerprint assessment, finding that temperatures in the present climate would be "unfamiliar" in the late 19$^{th}$ century, consistent with global warming.

In 1976, northwest Europe experienced a severe summer drought and heatwave, associated with a blocking pattern at the time. Baker et al., 2021 analysed temperature anomalies for a region covering England and Wales using long-term observations, AOGCM simulations representative of 1971 – 1980 and 2011 – 2020 conditions, and AGCM simulations with prescribed SSTs, sea ice, and GHG concentrations representative of 1976 and 2010 levels. They found that the probability of a summer at least as hot as 1976 increased significantly between the two simulated periods (by a factor of $4 - 28$ between the three methods), and that the joint probability of a hot summer following a dry winter and spring also increased by a factor of at least $13$.

The record European temperatures in 2014, as discussed previously, were also felt in England. An anthropogenic signal was found in record hot temperatures in central England since the 1990s (King, 2017). For this region, anthropogenic climate change increased the risk of warm years by a factor of at least $13$ (King et al., 2015) and of a new heat record by a factor $35$ (King, 2017), based on atmospheric and coupled model simulations respectively. More recently, provisional UK data indicated that 2022 may have been the warmest year so far recorded (Met Office, 2022). The UK also had its third warmest autumn on record in 2022 (BBC, 2022), which was preceded by an extreme summer, with the hottest day ever recorded in July in Lincolnshire, England (Andersson and Fauilkner, 2022). (Zachariah et al., 2022) focused on the summer heatwave in England, which impacted health and mortality, as well as bringing weather conditions conducive to the spread of wildfires (see Section 3.9). Using long-term observations and coupled models with current and preindustrial levels of GHGs, they found that anthropogenic climate change increased the risk of exceeding $40℃$ in England by at least $10$ times. However, they noted that their results were likely overly conservative, with models simulating a $2℃$ change in 1-day maximum temperature compared to $4℃$ in the observations.

Finally, there were two impact studies for heat-related mortality during the 2000s. The first, by Mitchell et al., 2016, looked at excess deaths in London, England during the 2003 European heatwave. They used a nested regional model in an AGCM with and without anthropogenic forcings to produce temperature and relative humidity data to calculate the number of heat-related deaths over June-August. They found that a 2003-like mortality event in London increased from a 1-in-7 to 1-in-2.5-year event due to anthropogenic influences, corresponding to a $20\%$ increased risk of heat-related mortality ($\mathrm{RR} = 1.25$), with $64$ out of the estimated $315$ deaths linked to the heatwave attributable to anthropogenic climate change. Then, Perkins-Kirkpatrick et al., 2022 performed an event and impact attribution for the 2006 heatwave in London to demonstrate why it "cannot be assumed that the anthropogenic signal behind the weather is equivalent to the signal behind the impact". They used atmosphere-only models with actual and counterfactual conditions and transfer functions between weather and impacts (as per Gasparrini et al., 2015). They reported that anthropogenic climate change increased the number of deaths associated with a 1-in-4-year heat event in London by $10$ deaths. Crucially, they noted the different $\mathrm{RR}$ results depending on whether the assessment was for temperature or mortality, and if using an event class or specific event approach. They found that $46–67\%$ of days with London temperatures above $26.6℃$ (corresponding to a mortality rate of at least $60$) were due to anthropogenic climate change ($\mathrm{RR} = 1.85 - 3$), whereas $37–50\%$ of deaths for the same mortality rate threshold were attributable ($\mathrm{RR} = 1.59 - 2$). This is in comparison to $17\%$, or around $10$ deaths out of $60$, found to be attributable when looking at the specific event. Another study into heat-related mortality found that mortality attributable fraction for high and extreme temperature with respect to minimum mortality temperature had increased in southeast England by $0.49$ in the previous two decades (Sahani et al., 2022).

#### *3.7.3.3 Scotland*

The event attribution study by Undorf et al., 2020 for the 2018 summer heatwave mentioned previously was conducted for Scotland specifically. AGCM simulations indicated that anthropogenic climate change increased the risk of high temperature extremes. For the hottest single day and warmest night, the best estimate $\mathrm{RR} = 1.2–2.4$ and $1.5– > 50$, respectively. Comparatively, Sahani et al., 2022 only found that mortality attributable fraction for high and extreme temperature with respect to minimum mortality temperature had reduced by 0.15 times in Aberdeenshire in the last two decades.

#### *3.7.3.4 Wales*

Studies on the May 2020 and summer 1976 heatwaves across western Europe discussed in above sections include coverage of Wales (Baker et al., 2021; Christidis and Stott, 2022a).

### 3.7.4 Summary & Gaps

Models have been found to generally overestimate temperatures in hot extremes (Wehrli et al., 2018). However, we found a high number of studies that utilised a range of methodologies and investigated heatwaves across spatial scales. All 33 studies here found an unequivocal anthropogenic role in UK heatwaves over the last couple of decades, and we are therefore most confident in attributing the change in heatwaves, of all hazards, to human influence. Nevertheless, there remains a gap in attributing heatwave impacts, with two of the three studies described above for events in London in the 2000s. Closing this gap is possible with current techniques, as demonstrated by Mitchell et al., 2016 and others.

## 3.8 Flood (snowmelt)

### 3.8.1 Background

A snowmelt flood is one that results from melting snowpack built up over winter, often due to a rapid rise in temperature. They can be exacerbated by heavy rainfall at the time of thawing, as well as frozen subsoil (Murray et al., 2021). Snowmelt floods can impact ecosystems and modulate river flows, with downstream impacts on water systems. With cold weather becoming milder with anthropogenic warming, and given the trend of increasingly sunny winters and springs (Kendon et al., 2022), snowmelt floods may become more common. However, the trend of fewer and less severe UK snowfall since the 1960s may offset this risk.

### 3.8.2 Key CIDs

The key CIDs are increasing mean and extreme high near-surface air temperatures, high sunshine and a change in mean snowfall. The impacts of snowmelt floods can be seen on daily timescales and over $10$s of km, given their contribution to river flows.

### 3.8.3 Literature

There are no D&A studies for snowmelt flood events specifically, however, two UK flooding studies which specifically considered snow processes are included in the following sections.

#### *3.8.3.1 UK*

Firstly, a study by Christidis et al., 2007 investigated the impact of human influence on the change in the growing season in the mid to late 20$^{th}$ century. They used a fingerprint approach with fully coupled models and found a detectable lengthening of the growing season on global to continental scales that could only be fully explained by taking human influence into account. They noted that associated earlier springs could affect the hydrological cycle, such as through earlier snowmelt generating earlier peak river flows. If water storage systems are unprepared, this could lead to summer water shortages. A later assessment of extreme UK winter sunshine using atmosphere-only models showed that human influence, particularly aerosols, increased the risk of extreme winter solar flux events by a factor of $3.6$, but noted that the record sunshine in winter 2014/15 occurred within a meteorological context atypical of sunny conditions (Christidis et al., 2016).

Kay et al., 2018 performed an event and impact attribution of the winter 2013/14 floods in the UK. They produced AGCM simulations with GHG and SST/SIC levels representative of the period, as well as a range of ensembles without anthropogenic GHGs and with different preindustrial-representative SST patterns. These fed into a hydrological model with a snow module. For daily peak flows, they found that past emissions increased the risk of peak river flows exceeding their chosen threshold in much of west England, Wales, and Scotland. They also found a notable contribution from snow in eastern Scotland and at longer durations, as well as an effect in southern England at shorter ($< 10$ days) durations, suggesting that snow changes moderate increases in shorter duration peak flows.

#### *3.8.3.2 England*

Prior to the above study, Kay et al., 2011 performed an event attribution for the unusual autumn/winter floods of 2000, in which snowmelt flooding was suggested as one contributor. They fed results from AGCMs with and without anthropogenic forcings into a hydrological model with a snow module for eight English river catchments. They found that emissions likely increased flooding risk over October–December, but that increased temperatures due to climate change were also likely to have reduced the risk of large snowmelt-induced floods.

### 3.8.4 Summary & Gaps

We looked at D&A studies for two UK winter flooding events which considered the role of snowmelt. Both used atmosphere-only models fed into a hydrological model and found that snowmelt exacerbated subsequent flooding risk, though significant impacts are likely felt in relatively few areas of the UK. Nevertheless, there seems to be a clear gap in understanding, with the lack of studies giving us low confidence in an already small attributable signal. The above studies could provide a method for further snowmelt flooding attribution going forward.

## 3.9 Wildfire

### 3.9.1 Background

Wildfires are any uncontrolled fire affecting a range of landscapes, from natural to residential (FAO, 2010; Murray et al., 2021). Wildfires can destroy ecosystems, working land and food systems, buildings, infrastructure, and impact human safety and air quality. With climate change, UK winters are projected to become milder and wetter, boosting plant growth and producing more fuel for wildfire, while projected increases in summer heatwaves and drought would dry out vegetation, making it more flammable. This suggests an increased risk of wildfires, and possibly an extended fire season.

### 3.9.2 Key CIDs

The key CIDs are extreme high near-surface air temperature with extreme low humidity and soil moisture, as well as lightning and extreme wind speed – both low and high, to respectively intensify and distribute the fire. Instantaneous lightning strikes and multi-day temperature, wind and humidity/moisture extremes can all cause impacts on sub-km scales in the UK.

### 3.9.3 Literature

There are no D&A studies specifically for UK wildfires. Here, we look at studies for three heatwaves across the UK. All of them noted the coincidence of drought conditions with high temperatures, together conducive to the wildfires that were observed at the time. Attribution of such related CIDs and hazards can provide insight into the anthropogenic influence on wildfire risk.

#### *3.9.3.1 UK*

One of the widely reported impacts of the 2018 UK summer heatwave was extensive and prolonged wildfires. McCarthy et al., 2019 used coupled models and found that the present-day risk of a summer temperature anomaly at or above that in 2018 was around $11 - 12\%$ – a factor of $30$ higher than for a world without anthropogenic GHG emissions.

#### *3.9.3.2 England*

The 2022 summer heatwave was linked to a high pressure system over Europe and warm air flowing from north Africa, and generated wildfires that caused damage throughout London most notably. Using long-term observations and coupled models with current and preindustrial levels of GHGs, Zachariah et al., 2022 found that anthropogenic climate change increased the risk of exceeding temperatures of $40℃$ in England by at least $10$ times. Model results also indicated a change in the 1-day maximum temperature intensity by $2℃$, compared to $4℃$ in observations. This event itself was very dry and followed a longer dry spell in the UK since the previous November. As a result of the higher background temperatures, soil moisture was also assumed to be particularly low due to higher evapotranspiration, which can itself enhance heatwave temperatures and contribute to wildfire risk.

#### *3.9.3.3 Scotland*

Undorf et al., 2020 also studied the 2018 summer heatwave but focused on Scotland, which saw widespread damage to newly planted trees and local biodiversity from wildfires, though was contained by sustained intervention. They used AGCMs with and without anthropogenic forcings and found that anthropogenic climate change had made such high temperature extremes more likely. For the hottest single day, they found $\mathrm{RR} = 1.2 - 2.4$, and for a 5-day mean $1.2 - 2.3$.

### 3.9.4 Summary & Gaps

The above attribution studies are for heatwave events across the UK, during which wildfires were also noted to occur. As a key CID, the anthropogenic effect on temperature extremes implies an effect on wildfires. Wildfires also depend on wind and soil moisture, though few drought attribution studies (see

Section 5.9) specifically link to associated wildfire risk. Given the damage caused by wildfires in recent years, the lack of D&A literature is a gap.

# 4 Wind-dominated Hazards

## 4.1 Extratropical Cyclone

### 4.1.1 Background

An extratropical cyclone is a low-pressure system that develops in latitudes outside of the tropics (Murray et al., 2021). Given their relatively large spatial extent, they can impact a variety of systems. Their frequency, location, and intensity are influenced by multiple properties of the global climate system, including atmospheric moisture and temperature gradients. With anthropogenic warming, atmospheric moisture content is expected to increase (Trenberth, 2011). Meanwhile the near-surface equator-to-pole temperature gradient is already decreasing due to Arctic amplification, and enhanced warming in the tropical upper troposphere and cooling in the near-polar stratosphere will lead to an increased gradient (Perlwitz, 2019). How these three factors will interact and modify extratropical cyclones, and the storm tracks as a whole, remains under investigation (Catto et al., 2019). Models suggest that over the UK, the number and intensity of the most extreme winter cyclones may marginally increase with anthropogenic climate change (Priestley and Catto, 2022), bringing heavier precipitation and stronger winds.

### 4.1.2 Key CIDs

The key CIDs are increasing mean and extreme high rainfall and wind speeds, as well as accompanying low atmospheric pressure. Low pressures can lead to impacts via downstream hazards (e.g., thunderstorms and storm surges), while multi-day build-up of rainfall, and sub-daily heavy rain and strong winds can themselves cause impacts over $10\text{s} - 100\text{s}$ of km.

### 4.1.3 Literature

There are no studies that attribute extratropical cyclone activity and impacts over the UK. In the following sections we predominantly present UK-specific D&A studies for rainfall events which noted the presence of storm systems.

#### *4.1.3.1 UK*

In an analysis of coupled model simulations, a distinct fingerprint of anthropogenic aerosol forcings was found in the weakening of the summer Eurasian subtropical westerly jet. This was suggested to have affected the number of extratropical cyclones and regional weather patterns. While the multi-model-mean showed higher summer precipitation over the UK in the aerosol simulations, when considered together with natural and GHG forcings, there was little change between 1979 and 2014 (Dong et al., 2022). Furthermore, dynamical relationships between circulation and weather are the subject of ongoing research.

The winter flooding of 2013/14 was associated with a record number of severe storms over the UK, driven by a particularly strong north Atlantic jet stream and unusually southerly storm track. Wild et al., 2015 sought to attribute the number of storms to human-influenced warming of the tropical west Pacific specifically. However, using long-term observations, they were unable to make this link. Later, event attributions for extreme high rainfall events that accompanied storm Desmond in December 2015 and storm Alex in October 2020 found that anthropogenic climate change had increased its risk ($\text{RR} = 2.6$ and $2.7$, respectively), suggesting some human influence on the cyclones that brought the extreme precipitation (Christidis et al., 2021a; Otto et al., 2018).

#### *4.1.3.2 England*

Schaller et al., 2016 examined flooding in winter 2013/14 in southern England using atmosphere-only models. They found that human-driven changes in atmospheric composition increased the risk of low pressure northwest of the UK and the number of days with zonal flow over the north Atlantic, and noted that SST response to anthropogenic forcing via a modification of the atmospheric meridional temperature gradient could impact mid-latitude cyclones.

### 4.1.4 Summary & Gaps

The D&A studies reviewed for extratropical cyclones focused on the associated CID of extreme high rainfall, rather than the driving cyclones themselves. However, attribution of a related CID, such as surface or sea level pressure (SLP), can also provide insight into the human influence on cyclones. There are D&A studies for sea level pressure change, though at global scales. Gillett et al., 2013 found that over Europe and the north Atlantic, the aerosol component of anthropogenic forcing likely drove most of the observed SLP reduction in the 20th century. However, as noted in Gillett and Stott, 2009, though the magnitude of the simulated and observed trends was typically larger in the mid-to-high latitudes, the variability there was also much larger, causing a much too low signal-to-noise to be accepted as attribution for the UK. They indicate that circulation patterns can be just as important a driver as anthropogenic forcings, though the contribution of such forcings to the development of these patterns has not been widely studied. Only Schaller et al., 2016 highlighted that changes in atmospheric composition due to human emissions affected the circulation associated with the particular rainfall event. Given that extratropical cyclones can lead to many downstream hazards, with subsequent impacts on all considered systems, the role of human influence in their behaviour is a gap in understanding.

## 4.2 Fog

### 4.2.1 Background

Fog refers to the suspension of $\mu$m-scale water droplets in the air, formed from water vapour in stable, humid near-surface air at the same temperature as its dew point. In the UK, freezing fog can be particularly dangerous during winter when temperatures are around $0°C$ and winds are low. Fog's primary effect is to reduce visibility (Murray et al., 2021), which impacts transport, most commonly in the winter. However, its formation requires specific wind, sunshine, and moisture levels, so its change with anthropogenic climate change is difficult to ascertain. Dew point temperature, for example, depends on surrounding air temperature and relative humidity, such that warming air containing fog might cause its evaporation. This suggests that fog may be less likely to form in a warmer climate.

### 4.2.2 Key CIDs

The key CIDs are high evaporation and humidity, and occurs on km and daily timescales.

### 4.2.3 Literature

There are currently no D&A studies that attribute changes in occurrence or impacts of fog.

### 4.2.4 Summary & Gaps

The lack of D&A studies for fog is partly due to a lack of records, as well as the fact that most climate models are unable to resolve the microphysical processes involved. However, this is not considered a gap, as fog does not pose a major risk in the UK.

## 4.3 Air Quality

### 4.3.1 Background

Air quality refers to the condition or level of pollution of air, considering chemical, physical, and biological characteristics (WHO, 2022). Rainfall can clear the air, but the interaction and contribution of other CIDs to air quality can depend on the pollutant itself. For inert, non-reactive chemicals, characteristics like particle size inform its impact, with smaller particles posing a higher risk to human health. While for reactive chemicals, such as nitrous oxide, which is emitted during agricultural and industrial activities, different meteorological drivers (e.g., wind, humidity, sunlight) can affect the rate of reaction and thus its behaviour when it comes into contact with humans. As such, poor air quality can have significant impacts on health, wellbeing, and mortality, as well as ecosystems and agriculture. As air quality is non-linearly impacted by a range of factors and CIDs, how its hazardous nature will change with anthropogenic warming remains unclear.

### 4.3.2 Key CIDs

The key CIDs are mean and extreme wind speed, near-surface air temperature, and rainfall, as well as evaporation, humidity, and soil moisture, with impacts across time and spatial scales.

### 4.3.3 Literature

There are currently no D&A studies that attribute changes in nor impacts of UK air quality to anthropogenic climate change.

### 4.3.4 Summary & Gaps

The lack of D&A studies is undoubtedly due to the highly complex nature of air quality, and the fact that its impacts have a significant human component. Mitchell et al., 2016 looked at extreme heat-related mortality but noted several factors that can modulate or contribute to such risk, including increased air pollution in heatwaves. Given the strong link between poor air quality and a range of health conditions, as well as the anthropogenically-driven increase in mean and extreme high temperatures that may contribute to such degradation, this is a knowledge gap.

## 4.4 Sting Jet

### 4.4.1 Background

A sting jet is the core of high winds ($\geq 100$ mph) that can form in a rapidly deepening area of low pressure and descend to the surface over several hours (Gentile and Gray, 2023; Met Office, 2019b). With anthropogenic climate change, frequency and intensity of extreme windstorms in the UK are expected to increase (Priestley and Catto, 2022). Sting jets develop in such cyclones, though are currently very rare. Nevertheless, a warmer and moister climate could produce more favourable conditions for sting jets, and they may become more frequent (Manning et al., 2022; Martínez-Alvarado et al., 2018). When they do occur, they impact mortality and various infrastructure systems.

### 4.4.2 Key CIDs

The key CID is extreme wind speed occurring at sub-daily and sub-km scales.

### 4.4.3 Literature

There are currently no D&A studies for sting jets.

### 4.4.4 Summary & Gaps

The closest proxy to sting jet attribution is that of cyclones or storms, however, these hazards are also lacking studies (see Sections 4.1 and 5.3, respectively). While warming is expected to affect storm intensity, the behaviour of associated or downstream hazards such as sting jets are more difficult to determine. Sting jets are historically rare in the UK's observational record, with some recent cyclones displaying indicators but lacking sufficient evidence to prove their presence. Furthermore, sting jets have only comparatively recently been assessed with high-resolution simulations, as most climate models use coarse grids to keep computation cheap. A limited number of studies using future climate scenarios suggest that they will become more common, but this is not well understood. Altogether, sting jets represent a gap in our D&A knowledge.

## 4.5 Storm Surge

### 4.5.1 Background

A storm surge refers to an unusually high water level caused by a meteorological disturbance, such as high winds or low atmospheric pressure (Murray et al., 2021). They are a combination of the piling up of water at the coast by winds and low pressure in a storm causing water levels to rise, which can be exacerbated by high tide. With rising sea levels and a potentially greater frequency of the most intense windstorms (Priestley and Catto, 2022), storm surges may become more common with anthropogenic climate change. They are hazardous in and of themselves, but particularly as part of compound flood events where sustained storm surges coincide with heavy rainfall and river discharge (Lyddon et al., 2023). Their impact can be seen on working land and seas, buildings, energy, transport, and mortality.

### 4.5.2 Key CIDs

Their associated extreme high wind speed, rising sea levels, and low surface pressure can cause impacts on sub-daily and km scales.

### 4.5.3 Literature

There are no D&A studies that attribute changes in occurrence or impacts of storm surges in the UK. Here, we look at studies for related CIDs and hazards, which provide some insight into the human influence on storm surges. Of the key CIDs, the low surface pressure is possibly hardest to attribute, especially given the relatively short time and spatial scale it must act over. For sea level rise, there are also no studies for the UK specifically, although Slangen et al., 2015 investigated 20th century trends in sea level rise globally using fully coupled models with a range of forcings. Discussed further in Section 5.1, this study found a significant human influence on sea level change. The extent to which general sea level rise may impact storm surges is likely to be dependent on the coincidence of stormy weather. However, high wind speeds associated with extratropical cyclones are similarly hard to attribute, with a distinct lack of D&A studies for cyclones noted in Section 4.1.

#### *4.5.3.1 England*

Schaller et al., 2016 studied winter 2013/14 flooding in southern England using AGCM simulations of the current and counterfactual climate. They found that human influence increased the risk of low pressure northwest of the UK and the number of days with zonal flow over the north Atlantic, with increased risk of high extreme rainfall as a result. Though not an attribution study for storm surges, they noted that the observed persistent atmospheric circulation pattern contributed to storm surges in large parts of southern England and Wales. With their model results indicating a role for anthropogenic climate change in generating such circulation, this also suggests an influence on storm surge occurrence.

### 4.5.4 Summary & Gaps

Given that storm surges are a well-known feature of UK coastlines, with documented impacts related to downstream hazards (see Sections 4.6 Flood (Coastal) and 4.7 Erosion (Coastal)) this is a gap in understanding.

## 4.6 Flood (Coastal)

### 4.6.1 Background

A coastal flood is one that results from, typically, a storm surge or high winds coinciding with high tide (Murray et al., 2021). They can significantly impact ecosystems, working land and seas, water, transport, and buildings. Given sea level rise with anthropogenic warming, and the potential for more storm surges that comes with a higher frequency of strong windstorms (Catto et al., 2019), coastal flooding may also be more common with climate change.

### 4.6.2 Key CIDs

The key CIDs are extreme high wind speed, rising sea levels, and low surface pressure, with impacts felt at sub-daily and km scales.

### 4.6.3 Literature

There are no D&A studies for coastal flooding and its impacts in the UK. Studying related CIDs and hazards can provide some insight into coastal flooding attribution. For example, coupled model studies indicated a significant human influence on both global mean expansion-driven sea level and regional dynamic sea level patterns in the 20th century (see Section 5.1 for detail) (Slangen et al., 2015). CIDs such as surface pressure and extreme wind can be linked to other related hazards, particularly storm surges and extratropical cyclones. But changes in these have also not been conclusively attributed to anthropogenic climate change. In Christidis and Stott, 2015, the atmospheric circulation that produced extreme rainfall in the UK winter of 2013/14 was linked to coastal flooding (and erosion) as a result of tidal surges triggered by a sequence of low pressure systems. However, this was not an explicit D&A study for a coastal flooding event.

### 4.6.4 Summary & Gaps

Given the wide range of potentially high-risk impacts associated with coastal flooding in the UK, the lack of studies on this hazard and its impacts is considered a gap.

## 4.7 Erosion (Coastal)

### 4.7.1 Background

Erosion is the removal of material from, in this case, coastal land, due to a combination of marine, fluvial, and landslide processes (Murray et al., 2021). This impacts ecosystems, transport, and buildings notably. The dependence on sea level and storm activity, which are both expected to rise with warming, means that coastal erosion may also be more common.

### 4.7.2 Key CIDs

The key CIDs are extreme high wind speed and rainfall, as well as high soil moisture and rising sea levels, with impacts occurring on daily to multi-day and km scales.

### 4.7.3 Literature

There are currently no D&A studies that attribute changes in nor impacts of UK coastal erosion.

### 4.7.4 Summary & Gaps

The closest proxy is attribution of coastal flooding (Section 4.6), which also lacks studies. Christidis and Stott, 2015 study referenced in Section 4.6.3 is one of the few to mention a link to coastal erosion, though ultimately does not find a significant human influence in the event. Given the damage that such events can cause (e.g., the destruction of the train tracks in Dawlish, England in 2014 (Dawson et al., 2016)), this is considered a gap. But as for river erosion (Section 5.7), model capability hinders progress.

# 5 Water-dominated Hazards

## 5.1 Sea Water Intrusion

### 5.1.1 Background

This is the process by which saltwater infiltrates and contaminates a body of rock and/or sediment holding fresh groundwater (i.e., an aquifer) (Murray et al., 2021). With anthropogenic warming-driven sea ice melt and ocean water expansion, the contamination of coastal aquifers is expected to accompany sea level rise. Intrusion of saltwater can impact ecosystems, agriculture and fisheries, and thus food and water systems.

### 5.1.2 Key CIDs

The key CIDs are increasing relative sea level and mean wind speed, occurring on multi-day timescales over 100s of km.

### 5.1.3 Literature

There are currently no D&A studies on sea water intrusion specifically. Here, we consider attribution of associated CID sea level, for which there have been studies on a global scale. Slangen et al. in 2014 and 2015 investigated sea level change in the 20$^{th}$ century (up to 2005) using coupled models with a range of forcings. The earlier study focused on the thermal expansion aspect, which accounts for approximately 40% of global mean sea level rise. They found that the trend could not be explained by natural variability alone; simulations with natural and anthropogenic forcings (aerosols, GHGs, ozone) showed a rise comparable with observations. The anthropogenic-only simulations contributed the most to this, suggesting that human influence had a significant impact on thermal expansion-driven sea level rise (Slangen et al., 2014). In the later study, this and regional dynamic sea level patterns (representing sea level change due to atmospheric and oceanic circulation changes and heat/salt redistribution) both showed responses to anthropogenic forcings that were significantly different from natural variability (Slangen et al., 2015).

### 5.1.4 Summary & Gaps

The above studies were conducted at a global scale, which is likely not representative of regional change. How this aspect of our water system has been impacted by anthropogenic warming is a knowledge gap.

## 5.2 Ocean Acidification

### 5.2.1 Background

Ocean acidification is the reduction of ocean pH over time. The primary mechanism is the dissolving of carbon dioxide ($CO_2$) from the atmosphere in sea water, driven by human emissions, though partly exacerbated by increased temperatures (Birchenough et al., 2017; Murray et al., 2021). There is high confidence that global-mean ocean pH has decreased since the preindustrial era as a result of the unprecedented rate of anthropogenic $CO_2$ uptake from the atmosphere, with the largest reduction in the northern North Atlantic (Hönisch et al., 2012; Rhein et al., 2013; Sabine et al., 2004). We expect responses in UK ecosystems in habitats and species (existing and new), with implications for conservation and fisheries (Birchenough et al., 2017).

### 5.2.2 Key CIDs

The key CIDs are increasing mean and extreme high ocean temperature, over multi-day timescales and 100s of km.

### 5.2.3 Literature

No formal attribution studies for acidification of seas surrounding the UK were found. There are, however, D&A studies for related ocean properties, such as heat content and salinity. Two such studies conducted a fingerprint analysis of trends from the mid-20$^{th}$ century onward and found the observed changes to be consistent with the changes expected due to anthropogenic forcings (Bilbao et al., 2019; Pierce et al., 2012). For ocean acidification itself, one study used an energy-balance carbon cycle model with and without anthropogenic forcings to quantify the contribution of $CO_2$ emissions from the 88 largest industrial carbon producers to global-scale ocean acidification. Their model results showed that approximately 55% of the decline in the surface ocean's pH over 1880 – 2015 could be traced to the emissions of those select companies (Licker et al., 2019).

### 5.2.4 Summary & Gaps

The process of ocean acidification is unequivocally dominated by $CO_2$ emissions. Thus, we can be confident that it is strongly attributable to human activities, though it is unclear how human-influenced changes in climate are modulating the process. One of the key impacts of this hazard is on ecosystems; for the UK, this would require adapting the so far open ocean-focused research to suit our coastal environments.

## 5.3 Thunderstorm

### 5.3.1 Background

Thunderstorms are a type of convective storm characterised by a sharp or rumbling sound (thunder) and lightning (WMO, 2020), typically accompanied by heavy precipitation (rain, snow, or hail) (Murray et al., 2021). As summer thunderstorms tend to follow hot, humid weather, they may become more common and intense in a warmer world with greater atmospheric moisture (Trenberth, 2011). The largest convective storms, which pose increased flood risks to Europe, are projected to increase in frequency and intensity (Chan et al., 2023). Such storms are associated with various other hazards besides floods (e.g., cyclones, hail, storm surges), and thus impact multiple systems.

### 5.3.2 Key CIDs

The key CIDs are extreme high rainfall, wind speed and humidity, as well as accompanying instantaneous, localised lightning and low atmospheric pressure system. The low pressures and sub-daily heavy rain and strong winds can cause impacts over 10s of km.

### 5.3.3 Literature

There are currently no D&A studies that attribute changes in characteristics, frequency, nor direct or indirect impacts of thunderstorms in the UK.

### 5.3.4 Summary & Gaps

There are no specific D&A studies for thunderstorms. Extreme rainfall events may be driven by or produce the conditions conducive to convective storms (e.g., warm and humid), and thus the closest related hazards are extratropical cyclones and pluvial flooding. The former also lacks D&A studies (see Section 4.1), and though the latter has more (see Section 5.4), the events may not necessarily be associated with a thunderstorm. This is therefore a knowledge gap, which depends on very high resolution convection-permitting models (CPMs) to be filled (Fowler et al., 2021a). Kahraman et al., 2022 used such an approach and projected a doubling of lightning counts in summer months with climate change, demonstrating a possible framework for attributing thunderstorms and related CIDs and hazards.

## 5.4 Flood (Pluvial)

### 5.4.1 Background

A pluvial flood results from high rainfall only (i.e., does not require overflow from a body of water) (Zurich, 2022). They can impact ecosystems, water systems, working land, buildings, and transport infrastructure. UK summers and winters have been on average wetter since the mid-20th century, with five of the ten wettest years since 1836 occurring this century (Kendon et al., 2022). Generally, a warmer atmosphere can hold more moisture (Trenberth, 2011). Observational trends and climate models show that short-duration (daily or shorter) extreme rainfall intensifies with the “Clausius-Clapeyron” rate, consistent with theory (Fischer and Knutti, 2016; Fowler et al., 2021a). Significant increases in extreme heavy rainfall have been found across the UK, emerging more quickly than changes in mean rainfall in some regions (Hawkins et al., 2020). There may also be a contribution from larger scale phenomena; atmospheric rivers are narrow bands in the atmosphere that carry moisture from the subtropics to midlatitudes and have been linked to extreme rainfall in the north and west of the UK particularly, with anthropogenic warming expected to make them more frequent (Hannaford, 2015).

### 5.4.2 Key CIDs

The key CIDs are extreme high and increasing mean rain and snowfall, acting on sub-daily to multi-day timescales, respectively, and over sub-km to 10s of km spatial scales.

### 5.4.3 Literature

A few studies reviewed looked at precipitation trends, and similarly few at specific extreme precipitation events, detailed as follows.

#### *5.4.3.1 UK*

Zhang et al., 2007 looked at general precipitation change over 1925 – 1999 using coupled models. Their best estimate suggested that anthropogenic forcing contributed around 50–85% of the observed 20th century trend in annual total land precipitation over northern midlatitudes, including the UK. A similar study found an anthropogenic contribution to European heavy precipitation using observations since the 1950s and coupled models (Dittus et al., 2016).

Wilcox et al., 2018 found a dominant influence of high-frequency variability on precipitation trends over Europe using long-term observations. They then focused on summer 2012, which saw widespread flooding in northern Europe, including the UK. Atmosphere-only simulations indicated $\mathrm{RR} = 0.95$, with the presence of a low pressure system that resulted in the heavy UK rainfall similarly attributed to natural variability rather than anthropogenic forcings. Very large ensembles of AGCM simulations with atmospheric gas constituents, SSTs and sea ice fractions representing 2012 and an analogous year in the preindustrial period also found minimal anthropogenic influence on this event (Sparrow et al., 2013). However, the authors did note that UK precipitation is sensitive to both the magnitude and pattern of warming in SSTs, which global warming can influence. Dong et al., 2013 investigated the role of certain circulation patterns using AGCMs with SSTs and sea ice representative of 1960 – 1990s and 2011 – 2012. North Atlantic SST anomalies were also found to be an important contributor to the atmospheric circulation over the north Atlantic and European region, with the south-shifted jet bringing cyclones across the UK into northern Europe, and with them, the observed extreme high rainfall in that period. However, Wilcox et al., 2018 reported no discernible contribution from anthropogenic-driven SST change to northern Europe precipitation in their atmospheric and coupled model simulations. Extreme high precipitation in May 2021 was similarly linked to persistent low pressure over the UK and a south-shifted jet stream. A study with coupled models with and without anthropogenic forcings (including emissions and land use change) found that the reference circulation pattern became more frequent with anthropogenic climate change, with its likelihood increasing by $18\%$ during the first half of the 21st century. They also noted that the persistent low pressure drove a considerable shift of the distribution toward higher rainfall, increasing the risk of extremes by approximately $3.5$. In their unconditional attribution, the best estimate $\mathrm{RR}$ was instead $1.5$ (Christidis and Stott, 2022b).

### 5.4.4 Summary & Gaps

The studies reviewed found notable contributions from coincident climatic conditions in specific events. Most found that these conditions occurred naturally, and one that there was an attributable contribution from anthropogenic climate change, though with large uncertainties. When considered with the related hazard of fluvial flooding, the size of the attributable signal in pluvial floods and our confidence in it are only moderate at most. Floods are the UK's most destructive hazard, impacting all considered systems. This uncertainty in the attribution is therefore a gap.

## 5.5 Flood (Fluvial)

### 5.5.1 Background

Fluvial floods result from a rise in the water level of a body of water. When the flow exceeds the capacity of a river/stream channel, it spills over the banks (natural or artificial) (Murray et al., 2021), with notable impacts on water, ecosystems, working land, buildings, and transport infrastructure. As with pluvial flooding, the expected change in fluvial floods with anthropogenic warming is generally an increase in intensity with the hydrological cycle, but depends strongly on season and region.

### 5.5.2 Key CIDs

The key CIDs are mean and extreme high rainfall and wind speed, high soil moisture and low surface pressure, leading to impacts on multi-day timescales, over sub-km to $10$s of km spatial scales.

### 5.5.3 Literature

The majority of the D&A studies reviewed focused on flooding events in the last two decades.

#### *5.5.3.1 UK*

Hannaford, 2015 conducted a trend attribution using observations from 1900 – 2012 to identify changes in UK river flow regimes. They found changes in river flows in the previous $40 - 50$ years, particularly around winter and in the western uplands, with mean outflows increasing over 1961 – 2011 across the UK. However, they found little indication of any longer-term trends in river flow (i.e., over the whole 20th century), nor any concurrent trends in flooding (or drought). Thus, they concluded there was no strong evidence for an anthropogenic influence on UK river flows, but that some changes resembled what might be expected with future warming (e.g., increases in winter flows). For the time around 2012 they were unable to attribute changes in river flow regimes to anthropogenic climate change due to the confounding effects of interdecadal variability particularly. In summer 2012 there was extreme high rainfall across Europe, with widespread fluvial flooding in the UK as a result. Wilcox et al., 2018 stated that if circulation analogous to that seen in summer 2012 generated the rainfall, then other, similar events could not be attributed to human-induced climate change. $\mathrm{RR}$ for the 2012 precipitation $= 0.95$ in northern Europe and they concluded that the presence of the low pressure system that brought the heavy rainfall was caused by natural variability.

The severe winter flooding of 2013/14 led to notable building damage and transport disruption, and was associated with a record number of severe storms driven by an unusually strong north Atlantic jet stream and southerly storm track (Christidis and Stott, 2015). However, anthropogenic climate change was only tenuously linked to the extreme rainfall at the time ($\mathrm{RR} = 1$ for December-February). Coupled models instead indicated that the event was made $8$ times more likely by the observed atmospheric circulation (Christidis and Stott, 2015). Vautard et al., 2016 took a slightly different approach for the same event. They used it as a case study to develop a methodology for D&A of externally forced dynamical and thermodynamical contributions to changes in the probabilities of extreme events. They used an AGCM and SSTs with and without an anthropogenic signal and found that, for 1-in-100-year events, human influence caused an average increase in the risk of extreme January precipitation in the south of the UK by approximately $40\%$, with thermodynamical changes explaining around two thirds of the change. Another study by Schaller et al., 2016 focused on England found the opposite, and is discussed more in Section 5.5.3.2. Taking a similar approach, Kay et al., 2018 conducted an event and impact attribution for England, Scotland and Wales, and the Thames catchment, respectively. They noted that although river flows were generally declining and below the seasonal average in early December 2013, they quickly increased in some areas with the mid-December stormy weather. Gauging stations indicated that floodplain inundations became more widespread around the end of 2013, and flows in many rivers in south, central and eastern England notably increased in January 2014. For their analysis, they ran AGCM simulations with GHG and SST/SIC levels representative of 2013/2014, and a range of ensembles without anthropogenic GHGs and with different preindustrial-representative SST patterns. The outputs were then fed into a hydrological model. For daily peak flows they found that past emissions increased the risk of peak river flows exceeding their chosen threshold in much of west England, Wales, and Scotland, with changes in snow processes affecting flows differently depending on the location (see Section 3.8).

Winter 2015/16 was also notably wet. In Christidis et al., 2018, this event was used to investigate how different methods can reach different conclusions (coupled vs. atmosphere-only models) and how preconditioning contributes to attribution, i.e., by simulating El Niño, QBO-W, etc. They found that anthropogenic forcings led to a modest increase in the likelihood of 1-in-10-year high extreme rainfall, by at least a factor of $1.5$– $2$. Otto et al., 2018 focused on the heavy precipitation brought to the north of the UK by storm Desmond that December. They used long-term observations with AGCMs and found that anthropogenic climate change increased the risk of such extreme daily and monthly rainfall events, though with some marginal and significant contribution from the coincident SST pattern, respectively. They found that anthropogenic climate change increased the likelihood of such a 1-in-5-year event (median $\mathrm{RR} = 2.6$), and that an upward trend in northern England/southern Scotland monthly precipitation of $20\%$ also increased the risk of extreme monthly total rainfall. While their study area was fairly large, they noted that precipitation in smaller areas, such as the basins of rivers that flooded, could be assumed to have a similar response, with implications for fluvial flooding. Storm Alex in October 2020 similarly brought strong winds and prolonged heavy rainfall, with 4-day accumulations reaching $150$ mm in some regions of the UK. Coupled model simulations indicated that human influence was not detectable in the historic mean trend in wettest day, but more in its variability. A specific extreme 3rd October rainfall event was found to be around $2.5$ times more likely with anthropogenic climate change (Christidis et al., 2021a).

#### 5.5.3.2 England

Flooding in autumn/winter 2000 was investigated in two event attribution studies. Pall et al., 2011 focused on England and Wales and used AGCM simulations representative of actual climatic conditions in 2000 and that which might have been without 20th century anthropogenic GHG emissions. They fed model outputs into a precipitation-runoff model to simulate severe daily river runoff as a proxy for flooding, and found that in $90\%$ of cases, anthropogenic forcings increased the risk of this event by more than $20\%$. In over $60\%$ of cases, the risk was increased by more than $90\%$. Kay et al., 2011 also used AGCMs representing an industrial and non-industrial climate to drive hydrological models of eight English catchments. They found that anthropogenic emissions likely increased the risk of flooding in October-December, with median $\mathrm{RR}$ values for 1-, 10- and 30-day durations $= 1.51$, $1.37$, and $1.2$, respectively. They also noted a likely reduction in the risk of large snowmelt-induced flood events as temperatures increased with climate change. Separately, they noted that for permeable catchments, groundwater stores did not deplete as much with anthropogenic influence.

Schaller et al., 2016 conducted an event attribution for the winter 2013/14 flooding, with focus on southern England, where there were large financial implications due to infrastructure and livelihood impacts. They used atmosphere-only models of the current and counterfactual climate and found that human influence increased the risk of low pressure northwest of the UK and the number of days with zonal flow over the north Atlantic. They also found that human influence on the climate (i.e., excluding interventions such as flood defences) increased the risk of high extreme precipitation in southern England. They fed modelled precipitation and temperature into a hydrological model of the Thames catchment and found that for a 1-in-100-year event, anthropogenic climate change had increased the modelled risk of 30-day peak river flows by around $21\%$, but daily peak flows by just $4\%$. For this event, they pointed to the quick succession of strong storms as the main driver of the Thames flooding, and noted that for this catchment, flooding is dominated by rainfall on shorter timescales, so had a marginal influence from human emissions. They then estimated that by the Thames, approximately $1{,}000$ more properties were at risk of flooding as a result of human-induced climate change, with potential losses of around $£24$MM. Kay et al., 2018 also looked at the Thames catchment and estimated that $457$ properties were at additional flooding risk that winter as a result of historical GHG emissions. As detailed previously, Otto et al., 2018 assessed heavy rainfall in 2015, finding that the likelihood of a very wet December in the north of England had increased by around $50 - 100\%$. Cotterill et al., 2021 similarly investigated the extreme high rainfall in autumn 2019 in England. They used long-term observations and AGCM simulations of 1900 and 2019 climate. They found that the frequency of observed extreme ($> 50$ mm) daily October-December UK-wide precipitation increased by $60\%$ between 1891-1920 and 1989-2018, which could imply anthropogenic climate change since the earlier period saw minimal human influence on climate. But contrary to the observations, the all and natural-only forcing simulations showed no appreciable difference, likely due to the model’s too coarse resolution.

#### 5.5.3.3 Scotland

Werritty, 2002 used observations for 1970 – 1996 to link precipitation trends in Scotland to climate change in a general sense. Their findings appeared to support claims that Scotland’s climate had become more variable toward the end of the 20th century, particularly in precipitation. They found that precipitation increases were concentrated in the winter half of the year, and were especially pronounced in north and west Scotland, with the drier east seeing a decrease in summer rainfall. They concluded that this was a “clear” climatic signal that could also be detected in river flow.

#### 5.5.3.4 Wales

See Section 5.5.3.2 for discussion of Pall et al., 2011, which investigated flooding in Wales in 2000.

### 5.5.4 Summary & Gaps

We found differing conclusions across the range of events, scales, and methodologies leveraged in the studies reviewed. Fluvial flooding has a significant impact on a growing number of communities around the UK, and as such, this is considered a gap. Closing this gap depends on the development of high quality data and modelling techniques, discussed further in Chapter 6.

## 5.6 Flash Flood

### 5.6.1 Background

A flash flood is one that is highly localised (within a few hundred $km^2$) and occurs over a few hours or less. They usually happen within $4-6$ hours of a causative event, which may be related to any one of the other types of flood (Murray et al., 2021). Where they follow from surface water or fluvial overflow, they occur multiple times a decade in various locations across the UK (Archer et al., 2019; Archer and Fowler, 2018). Due to their short-lived, high intensity nature, flash floods can be extremely destructive to human and ecosystems. Sub-daily and daily rainfall extremes are reported to be intensifying with warming at a rate consistent with the increase in atmospheric moisture, while in some regions, increases in short-duration intensities appear to be enhanced by convective cloud-related feedbacks (Fowler et al., 2021b, 2021a).

### 5.6.2 Key CIDs

The key CIDs are extreme high rainfall and soil moisture, acting on sub-daily timescales and over $\leq 10$s of km.

### 5.6.3 Literature

There are no existing D&A studies on UK flash flooding as per the above definition specifically, or its impacts. Since flash flooding can be considered a subset of the other types of flooding covered in this review, we refer you to Sections 4.6 Flood (Coastal), 5.4 Flood (Pluvial) and 5.5 Flood (Fluvial) for more information.

### 5.6.4 Summary & Gaps

There is a lack of long-term hourly-resolution observations, necessary to identify flash flood events (Darwish et al., 2021), as well as CPM counterfactual simulations, which hinders flash flooding attribution. While other type of floods may provide a proxy, the highly localised nature of flash floods and the potential for damage they present highlights this as a gap in our attribution knowledge.

## 5.7 Erosion (River)

### 5.7.1 Background

Erosion is the removal of material from, in this case, riverbanks, when the force of flowing water exceeds the resisting forces of soil and vegetation in the bank (Murray et al., 2021). Any change in river erosion with anthropogenic climate change depends on other flooding hazards. As for those, changes in river flows are season and region dependent, and can affect water and ecosystems, as well as buildings and transport infrastructure.

### 5.7.2 Key CIDs

The key CIDs are increasing mean and extreme high rainfall, and soil moisture, with their impacts occurring on multi-daily timescales and over $10$s of km.

### 5.7.3 Literature

As for coastal erosion, there are currently no D&A studies that attribute changes in behaviour nor impacts of UK river erosion.

### 5.7.4 Summary & Gaps

Given the damage that river erosion can cause, the lack of studies is considered a gap. Attribution related to fluvial flooding is the closest proxy. However, subsequent river erosion occurs on spatial scales much smaller than almost all models use and involves certain land characteristics or processes that are not traditionally accounted for.

## 5.8 Drought (Meteorological)

### 5.8.1 Background

A drought is a period of unusually dry weather. Meteorological drought refers to a lack of precipitation over a large area, persisting for over a month (Murray et al., 2021). In general, anthropogenic warming is expected to intensify the hydrological cycle (Trenberth, 2011), suggesting a change in the frequency of extended periods without rainfall. Nevertheless, meteorological drought can have significant impacts in all seasons, but most acutely in the summer and particularly on water and ecosystems, with downstream effects on working land and food systems.

### 5.8.2 Key CIDs

The key CIDs are decreasing mean and extreme low rainfall, occurring on multi-day timescales and over $10$s of km.

### 5.8.3 Literature

There are few D&A studies for meteorological drought that cover the UK.

#### *5.8.3.1 UK*

Gudmundsson and Seneviratne, 2016 investigated the change in 20-year return period low precipitation years as a proxy for meteorological drought. They used Europe-wide long-term observations for a fingerprint analysis using a two-step attribution procedure to link the observed increase in northern Europe precipitation over 1886 – 2005 to global mean warming, which has been attributed to human influence. They then used an ensemble of coupled model simulations with historic radiative forcing and with only natural forcing for decadal slices for the same period. The models corroborated their fingerprint analysis, and they confidently concluded that the changes in European drought risk, with a reduction in the UK, were attributable to anthropogenic climate change. Alternatively, in 1976, northwest Europe experienced a severe summer drought and heatwave coincident with an atmospheric blocking pattern. The event was preceded by a winter and spring with similarly low rainfall, associated with a north-shifted jet stream. Baker et al., 2021 analysed precipitation (and temperature) anomalies over England and Wales using long-term observations, AOGCM simulations representative of 1971-1980 and 2011-2020 conditions, and AGCM simulations with prescribed SSTs, sea ice, and GHG concentrations representative of 1976 and 2010 levels. They found that between the two simulated periods, the risk of a dry and hot summer increased by a factor of $10$ and $28$ for the respective modelling approaches. However, while the joint probability of a hot summer following a dry winter-spring had increased (by a factor of $13 - 33$), they did not find a significant change in the odds of a dry summer following a dry winter-spring (only a factor of $1.8 - 2.6$ change).

### 5.8.4 Summary & Gaps

Low precipitation can have serious implications for agricultural, water, and health systems. Wetting trends over large regions, such as northern Europe, may not necessarily be seen at more local scales (Baker et al., 2021), and hazards such as drought, in which each event has a distinctive signature, could benefit from event attributions that take a storyline approach. However, there are to-date insufficient studies to be able to properly assess how this hazard and its impacts have changed with human influence, which is a gap.

## 5.9 Drought (Hydrological)

### 5.9.1 Background

Hydrological drought refers to a period of low water supply and is often associated with precipitation shortfalls (Murray et al., 2021). As such, the expected change with anthropogenic warming depends on the region in question, though observed trends in UK drought do not seem to follow that of river flow regimes with climate change (Hannaford, 2015). Furthermore, research that suggests a trend towards earlier springs and later autumns indicates an impact on the timing of river flow peaks, with implications for water availability. Hydrological drought, by definition, impacts water systems, and consequently health, working land and food, and ecosystems.

### 5.9.2 Key CIDs

The key CIDs are decreasing mean and extreme low rainfall, humidity, and soil moisture, as well as increasing and extreme high near-surface air temperature and evaporation, generating impacts on multi-day timescales and over 10s of km.

### 5.9.3 Literature

There are no explicit D&A studies on hydrological drought in the UK, though those highlighted for meteorological drought (Section 5.8) are relevant.

#### *5.9.3.1 UK*

Building on those, Gudmundsson et al., 2017 used observations and coupled models with and without anthropogenic forcings to study pan-European river flow over 1956 – 2005. They found that it was likely that anthropogenic emissions had left a detectable fingerprint in renewable freshwater resources. Their models only recreated the observed European south-north dry-wet dichotomy if human emissions were accounted for, though the response was smaller than observed. They noted that this was consistent with changes in European meteorological drought risk (see Section 5.8), but that in northern Europe, including the UK, their regression-based approach did not detect anthropogenic influence in the observations if the effect of North Atlantic oscillation (NAO) was removed.

### 5.9.4 Summary & Gaps

Though meteorological drought may provide a proxy for hydrological drought, few D&A studies exist for either. And, while meteorological drought should only reflect climatic changes, hydrological drought can be impacted by human interventions as well as climate change (Gudmundsson and Seneviratne, 2016). Continental-scale studies tend to find a trend towards wetter years in northern Europe, suggesting a lower risk of drought (Gudmundsson et al., 2017). However, an average change may not reflect on seasonal and regional time scales, with Baker et al., 2021 finding that the risk of a hot and dry UK summer like 1976 has increased over time. Another similar, though not attribution, study found that the joint probability of long dry spells coinciding with high temperatures over a region including the UK increased for the period 1984 – 2013 compared to 1950 – 1979 (Manning et al., 2019). This was found to be driven by increasing temperatures in dry spells, which exacerbated evapotranspiration and soil moisture deficits. Though not fully attributing to anthropogenic climate change, this study infers a role in drought conditions. Given the potential large impacts on water, food, and health systems, we consider the overall uncertainty driven by a lack of attribution studies a gap.

## 5.10 Drought (Agricultural)

### 5.10.1 Background

Agricultural drought is a combination of factors including precipitation shortfall, soil water deficits, reduced groundwater, and increased evapotranspiration – all ultimately impacting agricultural production (Murray et al., 2021). It has significant impact on working land and food systems, and subsequently health, mortality, and communities. Given the interplay of temperature, sunshine, and precipitation that produces an agricultural drought, how it has and will change with anthropogenic climate change is unclear. Nevertheless, the risk of an extreme hot and dry UK summer is noted to have increased Baker et al., 2021, which would be consistent with global warming and the greater evaporation and lower soil moisture that greater temperatures and sunshine can bring.

### 5.10.2 Key CIDs

The key CIDs are decreasing mean and extreme low rainfall, humidity, and soil moisture, as well as increasing and extreme high near-surface air temperature and evaporation, with their effect felt on multi-day timescales and over 10s of km.

### 5.10.3 Literature

As for hydrological drought, agricultural drought is affected by rain shortfalls. Thus, studies such as Baker et al., 2021 and Gudmundsson and Seneviratne, 2016 can provide some indication of human influence (see Sections 5.8 and 5.9, respectively).

#### *5.10.3.1 UK*

Another study by Christidis et al., 2007 looked at global changes in growing season length; a temperature index that is commonly linked to agricultural, as well as terrestrial and freshwater ecosystem, impacts. This includes changes in production for certain crops with earlier springs, egg laying and breeding dates, but also potentially increases in insect and pest diseases. As noted in Section 3.8, earlier springs can also affect the hydrological cycle, such as via earlier snowmelt, and by prolonging and thus enhancing evapotranspiration. They found a detectable signal over 1950 – 1999 on both global and continental scales, which coupled model simulations showed could only be fully explained by including anthropogenic forcings. While they did not specify or quantify any downstream effects of this, their results indicate some contribution to agricultural drought. Furthermore, a global impact attribution study by Dasgupta and Robinson, 2022 linked human-driven heat and drought changes over 1981 – 2010 to food insecurity. Food security is simply about every person having unrestricted access to food that enables them to be healthy, so depends on both food availability and affordability. They used re-analysis data to find trends in heat stress via temperature anomalies and low precipitation (agricultural drought proxy), coupled model simulations with present day and counterfactual conditions, as well as over 400,000 responses to a food insecurity survey from the Food and Agriculture Organization. Overall, they found that temperature anomalies increased the probability of food insecurity, with the magnitude of this impact increasing over time. They inferred that climate change had worsened moderate and severe food security, across all regions. The UK data indicated a less severe impact compared to some other countries, consistent with other northern European countries; the incidence of moderate to severe food insecurity in Europe would have been 1.46 percentage-points lower without climate change, and for severe food insecurity, 0.19 percentage-points lower. They note that these differences in impacts are due to the differentiated impacts of the temperature anomaly on the two indicators of food insecurity (Dasgupta and Robinson, 2022).

### 5.10.4 Summary & Gaps

Dasgupta and Robinson, 2022 appears to be the only study for an agricultural drought impact with some data for the UK, though does not explicitly link to anthropogenic climate change. This is a gap, as differing localised and seasonal effects on agriculture can have widespread and long-lasting impacts throughout the UK and beyond.

## 5.11 Subsidence/Uplift

### 5.11.1 Background

Subsidence refers to the collapse or lowering of the ground, and uplift is the converse (BGS, 2020). Both can impact buildings, business, transport, energy, and telecoms and ICT infrastructure. The expected change with anthropogenic climate change is undetermined.

### 5.11.2 Key CIDs
The key CIDs are mean and extreme rain and snowfall, soil moisture, and rising sea levels, with impacts on multi-day and sub-km scales.

### 5.11.3 Literature
There are currently no D&A studies that attribute subsidence and uplift in the UK.

### 5.11.4 Summary & Gaps
The lack of D&A studies is not considered a gap, as neither subsidence nor uplift have historically posed a significant risk in the UK.

## 5.12 Landslide

### 5.12.1 Background
A landslide is the movement of a mass of rock, debris, or earth downslope due to downward-acting forces exceeding the strength of the material comprising the slope (USGS, 2021). They can damage buildings, transport, energy, and telecoms and ICT infrastructure. With anthropogenic climate change, we might expect the number of landslides to increase, such as by the coast (Lyddon et al., 2023) or in areas that become oversaturated by persistent or intense precipitation, in line with a change in flash flooding.

### 5.12.2 Key CIDs
The key CIDs are mean and extreme rain and snowfall, soil moisture, and rising sea levels, with impacts on multi-day and sub-km scales.

### 5.12.3 Literature
No D&A studies that attribute landslide occurrence or impacts were found for the UK.

### 5.12.4 Summary & Gaps
Landslides present a gap in D&A knowledge. The closest proxy is coastal and river erosion (Sections 4.7 and 5.7, respectively), but these also lack attribution studies. Additionally, landslides occur on spatial scales much smaller than used in most models, as well as depending on certain land characteristics or processes that are not typically accounted for.

## 5.13 Water Quality

### 5.13.1 Background
Water quality refers to the condition of water in relation to a specific use (i.e., drinking), and considers chemical, physical, and biological characteristics (NOAA, 2017). Water quality impacts water systems, and thus health, agriculture and food, and ecosystems. The wide range of types and geographical location of water bodies in the UK is likely to mean a similar range of changes in their quality with anthropogenic warming, which is at present not well understood.

### 5.13.2 Key CIDs
The key CIDs are mean and extreme rain and snowfall, and rising sea levels, with impacts across time and spatial scales.

### 5.13.3 Literature
There are currently no D&A studies for UK water quality. Here we look at studies for related CIDs or hazards.

##### 5.13.3.1 UK

Hannaford, 2015 conducted a trend attribution using observations over 1900 – 2012 to identify changes in river flow regimes. Detailed in Section 5.5.3, they found no strong evidence for anthropogenic warming influences on UK river flows.

##### 5.13.3.2 Wales

Durance and Ormerod, 2007 used observations from 1981 – 2005 to study impacts on stream ecosystems. With focus on the area around a reservoir in south Wales (Llyn Brianne), they reported that longer-term variations in temperature appeared to affect macroinvertebrates only in circumneutral ($pH = 7$) streams, with no long-term trend found in discharge at the study sites. This implied that climate change did not alter winter runoff. They also noted that their data suggested that the ecological consequence of general climate change for upland streams could be far reaching, with the biggest impact in the most species-rich locations.

#### 5.13.4 Summary & Gaps

As for air quality (Section 4.3), the lack of D&A studies is likely due to the range and complexity of drivers, as well as the non-linearity of CID interactions in contributing to water quality and its impacts. We consider this to be a knowledge gap, given the strong link between poor water quality and a range of health conditions and mortality.

# 6 Discussion & Conclusions

## 6.1 Summary of Findings

Overall, 67 D&A studies have been included as per our criteria in Table 1. This review utilises a good practice definition of D&A, requiring studies to apply statistical techniques to an observable quantity in order to show that any change is significantly inconsistent with noise. Many papers that claim to be using D&A are not this strict; for instance, by simply fitting a trend line to a timeseries. Figure 1 summarises our assessment of each of the 29 hazards using size of attributable signal and confidence, as well as noting direction of change. The latter refers to a hazard's change in risk or trend, depending on available studies. Confidence is informed by the number of and agreement between studies, where a high number of papers is $N \geq 10$ and low $N \leq 2$. Attribution signal is determined by $\mathrm{RR}$ or equivalent measure, with a strong $\mathrm{RR} \geq 4$ and weak $\mathrm{RR} = 1$: note that even a moderate $\mathrm{RR} = 2$ indicates a doubling of risk. Furthermore, our assessment is aggregated at the UK scale and across all months. This is key to our assessment of the drought-related hazards particularly; for example, agricultural drought may be of greatest concern in summer, but wetting trends are on average upward (i.e., risk of meteorological drought is overall lower) (Berry and Brown, 2021). The differing directions of change at an aggregated level explain our low to medium confidence, but this may have been higher if we were able to focus on the most relevant seasons and locations only. Some other results in Figure 1 worth noting include air and water quality, for which the increase in risk refers to an increase in incidences of poor quality. For ocean acidification, we recorded high confidence in a strong attributable signal as global studies have unequivocally linked ocean acidification to human $CO_2$ emissions in preceding decades (see Section 5.2 for details).

No UK-inclusive D&A literature was found for fourteen of the 29 hazards. The human influence on some of these could be inferred from associated hazards that did have studies (e.g., ice from cold waves), but confidence in attribution could not be high as a result. Even for the remaining fifteen hazards, few had studies specifically about the hazard itself nor its impacts, but rather relied on attribution of key CIDs or other related hazards (see for example: wildfire, extratropical cyclone, storm surge, sea water intrusion). Temperature-dominated hazards had the most studies (42), most of which investigated heatwaves, the single most-studied hazard for the UK. The next most-studied hazards were flood-related. Despite the relatively high number of studies, confidence was only medium at most, as there were often differing results across the studies. There was also a gap in D&A for compound or multivariate extremes; for example, anticyclonic systems link heatwaves and meteorological drought, with both ultimately having implications for working land and food systems, but lacking impact attribution. Overall, the most significant gaps raised by this review is the lack of nation-specific and impact attribution studies. A more detailed synthesis of our conclusions and recommendations are summarised further in this and the following section.

Heatwaves were the most studied hazard, with a unanimous consensus across all 33 studies of a strong attributable signal of human-induced climate change in their increased frequency and intensity over the last century. Though with fewer studies, the converse reduction in cold wave risk was similarly deemed strongly attributable. We had high confidence in the attribution of these hazards given the number of studies, as well as the fact that they utilised a variety of attribution approaches and that UK-focused studies corroborated larger-scale event and trend attributions. Nevertheless, these comparatively well-studied hazards still lacked impact attributions. The downward trend in cold extremes suggests an easing of pressure on related impact sectors, such as health, while the exacerbation of extreme heat events with warming has seen an increasing impact on health particularly, as well as a range of infrastructure sectors. The impacts of both cold and heatwaves are strongly related to societal measures to mitigate their effects, but without the studies to explicitly demonstrate it, the role of human-induced climate change compared to societal or infrastructural factors is unknown in these hazards' impacts. Flood-related studies also leveraged a range of attribution approaches across spatial scales (hemispheric to river catchments) and events. However, amongst the 19 pluvial and fluvial flooding studies for example, an anthropogenic signal was found in large-scale precipitation trends, but not necessarily in regional studies. There were also differing conclusions for individual compared to classes of flooding events. Similar results were found for drought-related hazards, though with far fewer studies (5). This uncertainty in flood and drought attribution meant our confidence in an anthropogenic signal could at most be moderate at an aggregated level, but could have been higher for certain seasons and regions. Storms and related hazards such as extratropical cyclones, sting jets, storm surges, and coastal flooding are in general poorly studied within D&A. They can have significant impacts on infrastructure, but it is difficult to model them in general and attribution studies are thus rare.

Most studies covered the UK generally (44). Of these, 18 focused on the UK alone, and the rest were European, Northern Hemisphere, or global studies which included data for the UK. As noted throughout the review, changes in CIDs and hazards at these large scales are not necessarily applicable for the UK and its individual nations. For temperature, anomalous UK heat is often found to co-occur with large-scale European extreme heat events associated with high-pressure systems (Yule et al., 2023), but precipitation is strongly influenced by local geography and European studies cannot necessarily explain more localised changes in the UK. The remaining 23 studies looked at one or more of the four UK nations specifically (shown in Figure 4). 20 studies focused on England, 4 on Scotland, 4 on Wales, and none on Northern Ireland. Within this, 2 investigated locations in both England and Scotland, and 3 England and Wales; the rest were solely about each individual nation. Greater geographic granularity could benefit devolved administrations, with attribution providing insight that could help prevent maladaptation. Improved understanding and quantification of the contribution of human activities to, for example, the changing frequency, magnitude, or extent of hazards and their related impacts can ensure that adaption and mitigation practises for existing and future climate change are most appropriately directed. Going beyond the devolved administrations, attribution studies could also focus on specific cities, which often have very different socioeconomic infrastructure from each other (e.g., flood drainage, number of hospitals). At the same time, many city councils have significant decision-making capabilities, funding, and green city strategies, and would therefore benefit from targeted attribution studies.

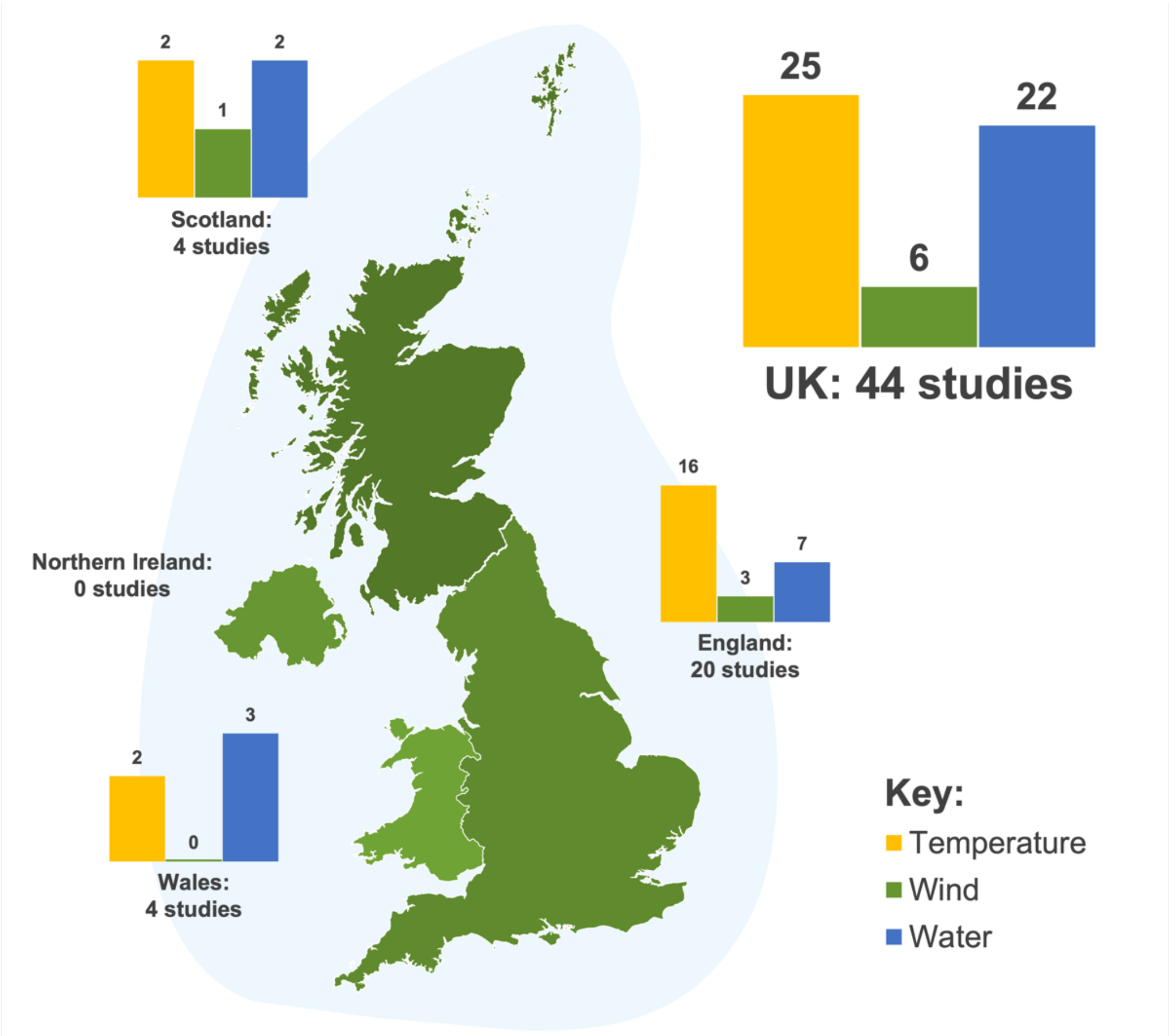

*Figure 4 – The breakdown of the studies reviewed by UK nation, with some covering multiple hazards and/or nations. Of the 44 studies covering the UK as a whole, 18 focused on the UK alone, while the rest covered the UK as part of a larger-scale investigation.*

## 6.2 Recommendations

Perhaps the most notable gap raised by the review is the lack of impact attribution studies. In CCRA3, the highest priority impact sectors for near-term adaptation point to certain hazards, with which a better understanding of drivers and impacts of existing changes could provide helpful insight to decision makers. Those impact systems include ecosystems, water, working land and seas, food, health, energy, built environment, and finance. For most of these, the most relevant hazards are flooding and drought, though heatwaves, and air and water quality are key to health and wellbeing. For a complete assessment of impacts on human-systems, it is just as important to consider societal as climatic change; the former can determine the ultimate impacts of natural hazards. For example, assessing the relative attribution of heat-related deaths to extreme temperatures vs. a heatwave plan, or of financial losses to extreme rainfall vs. flood defences, provides insight into climate change vs. adaptation time scales. For natural systems, impact attribution requires additional consideration of a wide range of anthropogenic influences, beyond climate change, on environmental factors that affect ecosystem vulnerability and disturbances (Bastos et al., 2023). As discussed in Section 1.2.3, this requires a multi-step approach. The Inter-Sectoral Impact Model Intercomparison Project (ISIMIP) is an initiative that investigates future impacts of climate change across affected sectors and spatial scales. Though typically projections-focused, this could provide a framework for impact attribution – the existing or relatively easily producible results of CID or hazard attributions can be fed into their specific impact models to estimate consequent changes on anything from energy to ecosystems. Other individual, non-attribution studies also suggest possible approaches to the attribution of impacts for many of the hazards covered in our review, with some detailed as follows.

Studies that have been able to quantify the role of anthropogenic climate change in flood-driven damage (Kay et al., 2018; Schaller et al., 2016) demonstrate possible approaches for attributing built environment-centred impacts of some of the UK's most destructive hazards. However, other impacts are less well studied; Koks et al., 2019 do not attribute to anthropogenic climate change, but link economic losses associated with power outages and business interruption to flooding in southeast England – an indication of what it is possible to cover in impact attribution. Another limitation is that comprehensive understanding of rainfall extremes and related flooding hazards requires high quality observational records and greater model resolution, which is currently much more computationally expensive (Kendon et al., 2018, 2014). Meanwhile, similarly few studies have attributed impacts of low flows; most tend to be forward looking. For example, one study that aimed to quantify the economic impacts of drought on the electricity sector under three future climate change scenarios used a methodology that could be utilised for an attribution of historical change (Byers et al., 2020). Such hazards may also benefit from so far underused techniques, such as the storyline approach to attribution, which considers the thermodynamical circumstances that have led to the occurrence of a particular extreme event. One storyline study found that intensity and duration of the 2010 – 2012 UK drought was highly influenced by the meteorological conditions preceding the drought, particularly in the north and at shorter timescales. They highlighted the importance of dry winters in the development of multi-year droughts, with autumn conditions a determinant of the timing of drought inception and winter conditions of drought length (Chan et al., 2021). Though this study goes on to place this event in the context of future warming, a similar approach could be utilised for historical climate change attribution.

For temperature, we found only a few studies attributing heat-related deaths in the UK to climate change (Mitchell et al., 2016; Perkins-Kirkpatrick et al., 2022). As for most impact systems, to have a full understanding of the impact of anthropogenic climate change on human health requires cross-disciplinary collaboration and integration of research ideas, techniques, and data (Mitchell, 2021). In general, studies of this kind either attribute the change in risk of the temperature extreme to anthropogenic influence and simply note impacts at the time (Ebi et al., 2020), or attribute the change in impacts to temperatures, but not necessarily to human drivers. For example, Sahani et al., 2022 investigated summer mortality in southeast England and Aberdeenshire, Scotland in relation to extreme high temperature and relative humidity. They found that in the last two decades, the attributable fraction of mortality to high and extreme temperature increased and reduced in the two regions, respectively. Their approach could be implemented in a multi-step mortality attribution study. Other research also indicates the importance of accounting for location, environment, biology, and infrastructure for comprehensive impact attribution, particularly when it comes to heat (de Schrijver et al., 2023; Iungman et al., 2023). The related hazard of wildfire was also notably understudied, despite growing incidence and impacts. Studies such as Liu et al., 2022 take a more global view but provide recommendations for future attribution efforts; to consider a variety of indices or metrics and understand the sensitivity of results to them, as well as better understanding the effect of climate change on the CIDs that produce fire weather (temperature, precipitation, wind speed, and humidity). Alternatively, cold has to-date driven domestic energy consumption, temperature-related mortality, and has wide-reaching implications for infrastructure throughout the UK. A recent study investigating the impact of future Arctic climate change on winter extremes conducted a return period analysis and found a strong multi-model-mean warming response of winter cold extremes, implying a lessening of adverse impacts on health, transport, and energy sectors over a region including the UK (Lo et al., 2023). A similar approach could be leveraged for attribution purposes. Furthermore, while it was noted that a sudden stratospheric warming (SSW) triggered the 2018 cold wave (Christidis and Stott, 2020), how such phenomena are affected by climate change remains the subject of debate and is complicated by the attribution of similar signed forcings, namely increased concentrations of CO2 and ozone depleting substances (Mitchell, 2016). Nevertheless, a recent study of SSW impacts on UK mortality could provide an example framework for a multi-step attribution of changes in cold-related mortality to anthropogenic climate change (Charlton-Perez et al., 2021).

For wind-dominated hazards, confidence was low due to a lack of studies, hindered by complexities involved in capturing such phenomena. A recent, though not D&A, study sought to identify the features of midlatitude cyclones that generate the strongest marine wind speeds, gusts, and wind wave heights, as well as the largest number of compound wind and wave hazard events in seas surrounding the UK (Gentile and Gray, 2023). This demonstrated that using data and conducting analysis on spatial scales small enough to resolve this behaviour could be very useful for attribution of cyclones and their impacts, especially storm surges and subsequent coastal flooding. This and other similar studies could therefore provide an example approach for future storm-related attribution (Gentile and Gray, 2023). Similarly, regional climate and convection-permitting models with grid spacing much smaller than typical GCMs can represent sting jets and cold conveyor belts at km-scale and thus better simulate UK windstorm intensity (Manning et al., 2023). Such models have been used for projecting future windstorms but could be used to close this gap in attribution knowledge for the UK.

The field of detection and attribution has rapidly expanded globally in the last few decades. Since emerging in the 2010s, impact attribution capabilities have also grown but remain a gap. This review highlights requirements for and opportunities to develop attribution science to meet the needs of the UK and makes us well placed to help fill the attribution knowledge gaps that exist not just for the UK, but globally.

# Glossary

| Acronym | Description |
|---|---|
| AGCM | Atmospheric GCM |
| AOGCM | Atmosphere-ocean GCM |
| AR | Assessment report |
| BL | Boundary layer |
| CCRA | Climate Change Risk Assessment |
| CET | Central England Temperature |
| CID | Climate impact driver |
| $CO_2$ | Carbon dioxide |
| CPM | Convection-permitting model |
| D&A | Detection and attribution |
| FAR | Fraction of attributable risk |
| GCM | General circulation model |
| GHG | Greenhouse gas |
| ICT | Information communication technology |
| IPCC | Intergovernmental Panel on Climate Change |
| ISIMIP | Inter-Sectoral Impact Model Intercomparison Project |
| NAO | North Atlantic oscillation |
| QBO | Quasi-biennial oscillation |
| RR | Risk ratio |
| SIC | Sea ice concentration |
| SST | Sea surface temperature |
| SSW | Sudden stratospheric warming |
| UK | United Kingdom |
| WG | Working Group |
| WWA | World Weather Attribution |

# Acknowledgements

Thank you to Lauren Brown, Joanne Godwin, and Alissa Haward for their support of this review, and Science Graphic Design for producing Figure 2. We have also engaged with a large number of people from across the D&A, impacts, and wider climate science community. Particular thanks go to Nicolas Bellouin, Jennifer Catto, Daniel Cotterill, Wolfgang Cramer, Hayley Fowler, Linda van Garderen, Nathan Gillett, Ed Hawkins, Abdullah Kahraman, Eunice Lo, Colin Manning, Matthew Priestley, Daniela Schmidt, Sebastian Sippel, Vikki Thompson, and Peter Watson.

# Bibliography

Allen, J.T., Giammanco, I.M., Kumjian, M.R., Jurgen Punge, H., Zhang, Q., Groenemeijer, P., Kunz, M., Ortega, K., 2019. Understanding Hail in the Earth System. Reviews of Geophysics 58, e2019RG000665. https://doi.org/10.1029/2019RG000665

Allen, M., 2003. Liability for climate change. Nature 421, 891–892. https://doi.org/10.1038/421891a

Allen, M.R., Tett, S.F.B., 1999. Checking for model consistency in optimal fingerprinting. Climate Dynamics 15, 419–434. https://doi.org/10.1007/s003820050291

AMS, 2012. Hail [WWW Document]. Glossary of Meteorology. URL https://glossary.ametsoc.org/wiki/Hail (accessed 6.1.23).

Andersson, J., Fauilkner, D., 2022. Heatwave: Fires blaze after UK passes 40C for first time. BBC News.

Archer, D., O'Donnell, G., Lamb, R., Warren, S., Fowler, H.J., 2019. Historical flash floods in England: New regional chronologies and database. Journal of Flood Risk Management 12, e12526. https://doi.org/10.1111/jfr3.12526

Archer, D. r., Fowler, H. j., 2018. Characterising flash flood response to intense rainfall and impacts using historical information and gauged data in Britain. Journal of Flood Risk Management 11, S121–S133. https://doi.org/10.1111/jfr3.12187

Baker, L., Shaffrey, L., Hawkins, E., 2021. Has the risk of a 1976 north-west European summer drought and heatwave event increased since the 1970s because of climate change? Quarterly Journal of the Royal Meteorological Society 147, 4143–4162. https://doi.org/10.1002/qj.4172

Bastos, A., Sippel, S., Frank, D., Mahecha, M.D., Zaehle, S., Zscheischler, J., Reichstein, M., 2023. A joint framework for studying compound ecoclimatic events. Nat Rev Earth Environ 4, 333–350. https://doi.org/10.1038/s43017-023-00410-3

BBC, 2022. Autumn 2022 is officially one of the warmest on record. BBC Weather.

Bellprat, O., Doblas-Reyes, F., 2016. Attribution of extreme weather and climate events overestimated by unreliable climate simulations. Geophysical Research Letters 43, 2158–2164. https://doi.org/10.1002/2015GL067189

Berry, P., Brown, I., 2021. National environment and assets, in: The Third UK Climate Change Risk Assessment Technical Report. Prepared for the Climate Change Committee.

Betts, R.A., Haward, A.B., Pearson, K.V., 2021. The Third UK Climate Change Risk Assessment. Prepared for the Climate Change Committee, London.

BGS, 2020. Swelling and shrinking soils. British Geological Survey. URL https://www.bgs.ac.uk/geology-projects/shallow-geohazards/clay-shrink-swell/ (accessed 6.1.23).

Bilbao, R.A.F., Gregory, J.M., Bouttes, N., Palmer, M.D., Stott, P., 2019. Attribution of ocean temperature change to anthropogenic and natural forcings using the temporal, vertical and geographical structure. Clim Dyn 53, 5389–5413. https://doi.org/10.1007/s00382-019-04910-1

Birchenough, S., Williamson, P., Turley, C., 2017. Future of the sea: ocean acidification. Government Office for Science.

Bloomfield, H., Thompson, V., 2023. Four ways winter heatwaves affect humans and nature [WWW Document]. The Conversation. URL http://theconversation.com/four-ways-winter-heatwaves-affect-humans-and-nature-197365 (accessed 6.2.23).

Brimelow, J.C., Burrows, W.R., Hanesiak, J.M., 2017. The changing hail threat over North America in response to anthropogenic climate change. Nature Clim Change 7, 516–522. https://doi.org/10.1038/nclimate3321

Byers, E.A., Coxon, G., Freer, J., Hall, J.W., 2020. Drought and climate change impacts on cooling water shortages and electricity prices in Great Britain. Nat Commun 11, 2239. https://doi.org/10.1038/s41467-020-16012-2

Catto, J.L., Ackerley, D., Booth, J.F., Champion, A.J., Colle, B.A., Pfahl, S., Pinto, J.G., Quinting, J.F., Seiler, C., 2019. The Future of Midlatitude Cyclones. Curr Clim Change Rep 5, 407–420. https://doi.org/10.1007/s40641-019-00149-4

Challinor, A., Benton, T.G., 2021. International dimensions. The Third UK Climate Change Risk Assessment Technical Report, RA Betts, AB Haward, and KV Pearson, eds.(Prepared for the Climate Change Committee).

Chan, S.C., Kendon, E.J., Fowler, H.J., Kahraman, A., Crook, J., Ban, N., Prein, A.F., 2023. Large-scale dynamics moderate impact-relevant changes to organised convective storms. Commun Earth Environ 4, 1–10. https://doi.org/10.1038/s43247-022-00669-2

Chan, W., Shepherd, T., Smith, K., Darch, G., Arnell, N., 2021. Storylines of UK drought based on the 2010-2012 event EGU21-1544. https://doi.org/10.5194/egusphere-egu21-1544

Charlton-Perez, A.J., Aldridge, R.W., Grams, C.M., Lee, R., 2019. Winter pressures on the UK health system dominated by the Greenland Blocking weather regime. Weather and Climate Extremes 25, 100218. https://doi.org/10.1016/j.wace.2019.100218

Charlton-Perez, A.J., Huang, W.T.K., Lee, S.H., 2021. Impact of sudden stratospheric warmings on United Kingdom mortality. Atmospheric Science Letters 22, e1013. https://doi.org/10.1002/asl.1013

Christiansen, B., 2015. The Role of the Selection Problem and Non-Gaussianity in Attribution of Single Events to Climate Change. Journal of Climate 28, 9873–9891. https://doi.org/10.1175/JCLI-D-15-0318.1

Christidis, N., Ciavarella, A., Stott, P.A., 2018. Different Ways of Framing Event Attribution Questions: The Example of Warm and Wet Winters in the United Kingdom Similar to 2015/16. Journal of Climate 31, 4827–4845. https://doi.org/10.1175/JCLI-D-17-0464.1

Christidis, N., McCarthy, M., Ciavarella, A., Stott, P.A., 2016. 10. Human Contribution to the Record Sunshine of Winter 2014/15 in the United Kingdom. Bulletin of the American Meteorological Society 97, S47–S50.

Christidis, N., McCarthy, M., Cotterill, D., Stott, P.A., 2021a. Record-breaking daily rainfall in the United Kingdom and the role of anthropogenic forcings. Atmospheric Science Letters 22, e1033. https://doi.org/10.1002/asl.1033

Christidis, N., McCarthy, M., Stott, P.A., 2021b. Recent decreases in domestic energy consumption in the United Kingdom attributed to human influence on the climate. Atmospheric Science Letters 22, e1062. https://doi.org/10.1002/asl.1062

Christidis, N., McCarthy, M., Stott, P.A., 2020. The increasing likelihood of temperatures above 30 to 40 °C in the United Kingdom. Nat Commun 11, 3093. https://doi.org/10.1038/s41467-020-16834-0

Christidis, N., Stott, P.A., 2022a. Anthropogenic Climate Change and the Record-High Temperature of May 2020 in Western Europe. Bulletin of the American Meteorological Society 103, S33–S37.

Christidis, N., Stott, P.A., 2022b. The Extremely Wet May of 2021 in the United Kingdom. Bulletin of the American Meteorological Society 103, E2912–E2916. https://doi.org/10.1175/BAMS-D-22-0108.1

Christidis, N., Stott, P.A., 2021. Extremely warm days in the United Kingdom in winter 2018/19. Bulletin of the American Meteorological Society 102, S39–S44.

Christidis, N., Stott, P.A., 2020. The Extremely Cold Start of the Spring of 2018 in the United Kingdom. Bulletin of the American Meteorological Society 101, S23–S28.

Christidis, N., Stott, P.A., 2015. 10. Extreme Rainfall in the United Kingdom During Winter 2013/14: The Role of Atmospheric Circulation and Climate Change. Bulletin of the American Meteorological Society 96, S46–S50.

Christidis, N., Stott, P.A., 2012. Lengthened odds of the cold UK winter of 2010/11 attributable to human influence. Bulletin of the American Meteorological Society 93, 1060–1062.

Christidis, N., Stott, P.A., Brown, S., Karoly, D.J., Caesar, J., 2007. Human Contribution to the Lengthening of the Growing Season during 1950–99. Journal of Climate 20, 5441–5454. https://doi.org/10.1175/2007JCLI1568.1

Christidis, N., Stott, P.A., Ciavarella, A., 2014. The effect of anthropogenic climate change on the cold spring of 2013 in the United Kingdom. Bulletin of the American Meteorological Society 95, S79–S82.

Christidis, N., Stott, P.A., McCarthy, M., 2023. An attribution study of the UK mean temperature in year 2022.

Christidis, N., Stott, P.A., Zwiers, F.W., Shiogama, H., Nozawa, T., 2012. The contribution of anthropogenic forcings to regional changes in temperature during the last decade. Clim Dyn 39, 1259–1274. https://doi.org/10.1007/s00382-011-1184-0

Christidis, N., Stott, P.A., Zwiers, F.W., Shiogama, H., Nozawa, T., 2010. Probabilistic estimates of recent changes in temperature: a multi-scale attribution analysis. Clim Dyn 34, 1139–1156. https://doi.org/10.1007/s00382-009-0615-7

Cotterill, D., Stott, P., Christidis, N., Kendon, E., 2021. Increase in the frequency of extreme daily precipitation in the United Kingdom in autumn. Weather and Climate Extremes 33, 100340. https://doi.org/10.1016/j.wace.2021.100340

Darwish, M.M., Tye, M.R., Prein, A.F., Fowler, H.J., Blenkinsop, S., Dale, M., Faulkner, D., 2021. New hourly extreme precipitation regions and regional annual probability estimates for the UK. International Journal of Climatology 41, 582–600. https://doi.org/10.1002/joc.6639

Dasgupta, S., Robinson, E.J., 2022. Attributing changes in food insecurity to a changing climate. Scientific Reports 12, 1–11.

Dawson, D., Shaw, J., Roland Gehrels, W., 2016. Sea-level rise impacts on transport infrastructure: The notorious case of the coastal railway line at Dawlish, England. Journal of Transport Geography 51, 97–109. https://doi.org/10.1016/j.jtrangeo.2015.11.009

de Schrijver, E., Royé, D., Gasparrini, A., Franco, O.H., Vicedo-Cabrera, A.M., 2023. Exploring vulnerability to heat and cold across urban and rural populations in Switzerland. Environ. Res.: Health 1, 025003. https://doi.org/10.1088/2752-5309/acab78

Dessens, J., Berthet, C., Sanchez, J.L., 2015. Change in hailstone size distributions with an increase in the melting level height. Atmospheric Research 158–159, 245–253. https://doi.org/10.1016/j.atmosres.2014.07.004

Dittus, A.J., Karoly, D.J., Lewis, S.C., Alexander, L.V., Donat, M.G., 2016. A Multiregion Model Evaluation and Attribution Study of Historical Changes in the Area Affected by Temperature and Precipitation Extremes. Journal of Climate 29, 8285–8299. https://doi.org/10.1175/JCLI-D-16-0164.1

Dong, B., Sutton, R., Shaffrey, L., 2014. The 2013 hot, dry summer in Western Europe. Bulletin of the American Meteorological Society 95, S61–S66.

Dong, B., Sutton, R., Woollings, T., 2013. The extreme European summer 2012. Bulletin of the American Meteorological Society 94, s28–s32.

Dong, B., Sutton, R.T., Shaffrey, L., Harvey, B., 2022. Recent decadal weakening of the summer Eurasian westerly jet attributable to anthropogenic aerosol emissions. Nat Commun 13, 1148. https://doi.org/10.1038/s41467-022-28816-5

Dooks, T., 2023. Progress in adapting to climate change 2023 Report to Parliament. CCC.

Durance, I., Ormerod, S.J., 2007. Climate change effects on upland stream macroinvertebrates over a 25-year period. Global Change Biology 13, 942–957. https://doi.org/10.1111/j.1365-2486.2007.01340.x

Ebi, K.L., Åström, C., Boyer, C.J., Harrington, L.J., Hess, J.J., Honda, Y., Kazura, E., Stuart-Smith, R.F., Otto, F.E.L., 2020. Using Detection And Attribution To Quantify How Climate Change Is Affecting Health. Health Affairs 39, 2168–2174. https://doi.org/10.1377/hlthaff.2020.01004

Ebi, K.L., Ogden, N.H., Semenza, J.C., Woodward, A., 2017. Detecting and Attributing Health Burdens to Climate Change. Environmental Health Perspectives 125, 085004. https://doi.org/10.1289/EHP1509

Eyring, V., Gillett, N.P., Achutarao, K., Barimalala, R., Barreiro Parrillo, M., Bellouin, N., Cassou, C., Durack, P., Kosaka, Y., McGregor, S., Min, S., Morgenstern, O., Sun, Y., 2021. Human Influence on the Climate System. In Climate Change 2021: The Physical Science Basis. Contribution of Working Group I to the Sixth Assessment Report of the Intergovernmental Panel on Climate Change, in: Masson-Delmotte, V., Zhai, V., Pirani, A., Conners, S.L., Péan, C., Berger, S., Caud, N., Chen, Y., Goldfarb, L., Gomis, M.I., Huang, M., Leitzell, K., Lonnoy, E., Matthews, J.B.R., Maycock, T.K., Waterfield, T., Yelekçi, O., Yu, R., Zhou, B. (Eds.), IPCC Sixth Assessment Report. Cambridge University Press.

FAO, 2010. Fire management | FAO TERM PORTAL | Food and Agriculture Organization of the United Nations [WWW Document]. Food and Agriculture Organization of the United Nations. URL https://www.fao.org/faoterm/collection/fire-management/en/ (accessed 6.1.23).

Fischer, E.M., Knutti, R., 2016. Observed heavy precipitation increase confirms theory and early models. Nature Clim Change 6, 986–991. https://doi.org/10.1038/nclimate3110

Fowler, H.J., Ali, H., Allan, R.P., Ban, N., Barbero, R., Berg, P., Blenkinsop, S., Cabi, N.S., Chan, S., Dale, M., Dunn, R.J.H., Ekström, M., Evans, J.P., Fosser, G., Golding, B., Guerreiro, S.B., Hegerl, G.C., Kahraman, A., Kendon, E.J., Lenderink, G., Lewis, E., Li, X., O'Gorman, P.A., Orr, H.G., Peat, K.L., Prein, A.F., Pritchard, D., Schär, C., Sharma, A., Stott, P.A., Villalobos-Herrera, R., Villarini, G., Wasko, C., Wehner, M.F., Westra, S., Whitford, A., 2021a. Towards advancing scientific knowledge of climate change impacts on short-duration rainfall extremes. Philosophical Transactions of the Royal Society A: Mathematical, Physical and Engineering Sciences 379, 20190542. https://doi.org/10.1098/rsta.2019.0542

Fowler, H.J., Lenderink, G., Prein, A.F., Westra, S., Allan, R.P., Ban, N., Barbero, R., Berg, P., Blenkinsop, S., Do, H.X., Guerreiro, S., Haerter, J.O., Kendon, E.J., Lewis, E., Schaer, C., Sharma, A., Villarini, G., Wasko, C., Zhang, X., 2021b. Anthropogenic intensification of short-duration rainfall extremes. Nat Rev Earth Environ 2, 107–122. https://doi.org/10.1038/s43017-020-00128-6

Gasparrini, A., Guo, Y., Hashizume, M., Lavigne, E., Zanobetti, A., Schwartz, J., Tobias, A., Tong, S., Rocklöv, J., Forsberg, B., Leone, M., De Sario, M., Bell, M.L., Guo, Y.-L.L., Wu, C., Kan, H., Yi, S.-M., de Sousa Zanotti Stagliorio Coelho, M., Saldiva, P.H.N., Honda, Y., Kim, H., Armstrong, B., 2015. Mortality risk attributable to high and low ambient temperature: a multicountry observational study. The Lancet 386, 369–375. https://doi.org/10.1016/S0140-6736(14)62114-0

Gasparrini, A., Masselot, P., Scortichini, M., Schneider, R., Mistry, M.N., Sera, F., Macintyre, H.L., Phalkey, R., Vicedo-Cabrera, A.M., 2022. Small-area assessment of temperature-related mortality risks in England and Wales: a case time series analysis. The Lancet Planetary Health 6, e557–e564. https://doi.org/10.1016/S2542-5196(22)00138-3

Gentile, E.S., Gray, S.L., 2023. Attribution of observed extreme marine wind speeds and associated hazards to midlatitude cyclone conveyor belt jets near the British Isles. International Journal of Climatology 43, 2735–2753. https://doi.org/10.1002/joc.7999

Gillett, N.P., Fyfe, J.C., Parker, D.E., 2013. Attribution of observed sea level pressure trends to greenhouse gas, aerosol, and ozone changes. Geophysical Research Letters 40, 2302–2306. https://doi.org/10.1002/grl.50500

Gillett, N.P., Stott, P.A., 2009. Attribution of anthropogenic influence on seasonal sea level pressure. Geophysical Research Letters 36. https://doi.org/10.1029/2009GL041269

Gudmundsson, L., Seneviratne, S.I., 2016. Anthropogenic climate change affects meteorological drought risk in Europe. Environ. Res. Lett. 11, 044005. https://doi.org/10.1088/1748-9326/11/4/044005

Gudmundsson, L., Seneviratne, S.I., Zhang, X., 2017. Anthropogenic climate change detected in European renewable freshwater resources. Nature Clim Change 7, 813–816. https://doi.org/10.1038/nclimate3416

Hannaford, J., 2015. Climate-driven changes in UK river flows: A review of the evidence. Progress in Physical Geography 39, 29–48.

Hansen, G., Stone, D., Auffhammer, M., Huggel, C., Cramer, W., 2016. Linking local impacts to changes in climate: a guide to attribution. Reg Environ Change 16, 527–541. https://doi.org/10.1007/s10113-015-0760-y

Hasselmann, K., 1993. Optimal Fingerprints for the Detection of Time-dependent Climate Change. Journal of Climate 6, 1957–1971. https://doi.org/10.1175/1520-0442(1993)006<1957:OFFTDO>2.0.CO;2

Hawkins, E., Frame, D., Harrington, L., Joshi, M., King, A., Rojas, M., Sutton, R., 2020. Observed Emergence of the Climate Change Signal: From the Familiar to the Unknown. Geophysical Research Letters 47, e2019GL086259. https://doi.org/10.1029/2019GL086259

Hegerl, G., Zwiers, F., 2011. Use of models in detection and attribution of climate change. WIREs Climate Change 2, 570–591. https://doi.org/10.1002/wcc.121

Hegerl, G.C., Hoegh-Guldberg, O., Casassa, G., Hoerling, M., Kovats, S., Parmesan, C., Pierce, D., Stott, P., 2010. Good practice guidance paper on detection and attribution related to anthropogenic climate change.

Hönisch, B., Ridgwell, A., Schmidt, D.N., Thomas, E., Gibbs, S.J., Sluijs, A., Zeebe, R., Kump, L., Martindale, R.C., Greene, S.E., Kiessling, W., Ries, J., Zachos, J.C., Royer, D.L., Barker, S., Marchitto, T.M., Moyer, R., Pelejero, C., Ziveri, P., Foster, G.L., Williams, B., 2012. The Geological Record of Ocean Acidification. Science 335, 1058–1063. https://doi.org/10.1126/science.1208277

Hope, P., Cramer, W., Flato, G., Frieler, K., Gillett, N.P., Huggel, C., Minx, J., Otto, F, Parmesan, C., Rogelj, J., Rojas, M., Seneviratne, S.I., Slangen, A.B.A., Stone, D., Terray, L., van Aalst, M., Vautard, R., Zhang, X., 2022. Cross-Working Group Box | Attribution.

IPCC, 2021. Summary for Policymakers Headline Statements [WWW Document]. IPCC Sixth Assessment Report Working Group 1: The Physical Science Basis. URL https://www.ipcc.ch/report/ar6/wg1/resources/spm-headline-statements/ (accessed 6.1.23).

Iungman, T., Cirach, M., Marando, F., Barboza, E.P., Khomenko, S., Masselot, P., Quijal-Zamorano, M., Mueller, N., Gasparrini, A., Urquiza, J., Heris, M., Thondoo, M., Nieuwenhuijsen, M., 2023. Cooling cities through urban green infrastructure: a health impact assessment of European cities. The Lancet 401, 577–589. https://doi.org/10.1016/S0140-6736(22)02585-5

James, R., Otto, F., Parker, H., Boyd, E., Cornforth, R., Mitchell, D., Allen, M., 2014. Characterizing loss and damage from climate change. Nature Clim Change 4, 938–939. https://doi.org/10.1038/nclimate2411

James, R.A., Jones, R.G., Boyd, E., Young, H.R., Otto, F.E.L., Huggel, C., Fuglestvedt, J.S., 2019. Attribution: How Is It Relevant for Loss and Damage Policy and Practice?, in: Mechler, R., Bouwer, L.M., Schinko, T., Surminski, S., Linnerooth-Bayer, J. (Eds.), Loss and Damage from Climate Change: Concepts, Methods and Policy Options, Climate Risk Management, Policy and Governance. Springer International Publishing, Cham, pp. 113–154. https://doi.org/10.1007/978-3-319-72026-5_5

Jaroszweski, D., Wood, R., Chapman, L., 2021. Infrastructure. Technical Report of the Third UK Climate Change Risk Assessment, Technical Report of the Third UKL Climate Change Risk Assessment Chapter 4.

Jones, G.S., Stott, P.A., Christidis, N., 2008. Human contribution to rapidly increasing frequency of very warm Northern Hemisphere summers. Journal of Geophysical Research: Atmospheres 113. https://doi.org/10.1029/2007JD008914

Kahraman, A., Kendon, E.J., Fowler, H.J., Wilkinson, J.M., 2022. Contrasting future lightning stories across Europe. Environ. Res. Lett. 17, 114023. https://doi.org/10.1088/1748-9326/ac9b78

Kaminski, I., 2022. How scientists are helping sue over climate change. The Lancet Planetary Health 6, e386–e387. https://doi.org/10.1016/S2542-5196(22)00098-5

Karoly, D.J., Stott, P.A., 2006. Anthropogenic warming of central England temperature. Atmospheric Science Letters 7, 81–85. https://doi.org/10.1002/asl.136

Kay, A.L., Booth, N., Lamb, R., Raven, E., Schaller, N., Sparrow, S., 2018. Flood event attribution and damage estimation using national-scale grid-based modelling: Winter 2013/2014 in Great Britain. International Journal of Climatology 38, 5205–5219. https://doi.org/10.1002/joc.5721

Kay, A.L., Crooks, S.M., Pall, P., Stone, D.A., 2011. Attribution of Autumn/Winter 2000 flood risk in England to anthropogenic climate change: A catchment-based study. Journal of Hydrology 406, 97–112. https://doi.org/10.1016/j.jhydrol.2011.06.006

Kay, G., Dunstone, N., Smith, D., Dunbar, T., Eade, R., Scaife, A., 2020. Current likelihood and dynamics of hot summers in the UK. Environ. Res. Lett. 15, 094099. https://doi.org/10.1088/1748-9326/abab32

Kendon, E.J., Blenkinsop, S., Fowler, H.J., 2018. When Will We Detect Changes in Short-Duration Precipitation Extremes? Journal of Climate 31, 2945–2964. https://doi.org/10.1175/JCLI-D-17-0435.1

Kendon, E.J., Roberts, N.M., Fowler, H.J., Roberts, M.J., Chan, S.C., Senior, C.A., 2014. Heavier summer downpours with climate change revealed by weather forecast resolution model. Nature Clim Change 4, 570–576. https://doi.org/10.1038/nclimate2258

Kendon, M., McCarthy, M., Jevrejeva, S., Matthews, A., Sparks, T., Garforth, J., Kennedy, J., 2022. State of the UK Climate 2021. International Journal of Climatology 42, 1–80. https://doi.org/10.1002/joc.7787

Kendon, M., Sexton, D., McCarthy, M., 2020. A temperature of 20°C in the UK winter: a sign of the future? Weather 75, 318–324. https://doi.org/10.1002/wea.3811

King, A.D., 2017. Attributing Changing Rates of Temperature Record Breaking to Anthropogenic Influences. Earth's Future 5, 1156–1168. https://doi.org/10.1002/2017EF000611

King, A.D., Oldenborgh, G.J. van, Karoly, D.J., Lewis, S.C., Cullen, H., 2015. Attribution of the record high Central England temperature of 2014 to anthropogenic influences. Environ. Res. Lett. 10, 054002. https://doi.org/10.1088/1748-9326/10/5/054002

Koks, E., Pant, R., Thacker, S., Hall, J.W., 2019. Understanding Business Disruption and Economic Losses Due to Electricity Failures and Flooding. Int J Disaster Risk Sci 10, 421–438. https://doi.org/10.1007/s13753-019-00236-y

Kovats, S., Brisley, R., 2021. Health, communities and the built environment. Technical Report of the Third UK Climate Change Risk Assessment, Technical Report of the Third UKL Climate Change Risk Assessment Chapter 5.

Lane, R.A., Coxon, G., Freer, J., Seibert, J., Wagener, T., 2022. A large-sample investigation into uncertain climate change impacts on high flows across Great Britain. Hydrology and Earth System Sciences 26, 5535–5554. https://doi.org/10.5194/hess-26-5535-2022

Leach, N.J., Li, S., Sparrow, S., van Oldenborgh, G.J., Lott, F.C., Weisheimer, A., Allen, M.R., 2020. Anthropogenic Influence on the 2018 Summer Warm Spell in Europe: The Impact of Different Spatio-Temporal Scales. Bulletin of the American Meteorological Society 101, S41–S46.

Leach, N.J., Weisheimer, A., Allen, M.R., Palmer, T., 2021. Forecast-based attribution of a winter heatwave within the limit of predictability. Proceedings of the National Academy of Sciences 118, e2112087118. https://doi.org/10.1073/pnas.2112087118

Licker, R., Ekwurzel, B., Doney, S.C., Cooley, S.R., Lima, I.D., Heede, R., Frumhoff, P.C., 2019. Attributing ocean acidification to major carbon producers. Environ. Res. Lett. 14, 124060. https://doi.org/10.1088/1748-9326/ab5abc

Liu, Z., Eden, J.M., Dieppois, B., Blackett, M., 2022. A global view of observed changes in fire weather extremes: uncertainties and attribution to climate change. Climatic Change 173, 14. https://doi.org/10.1007/s10584-022-03409-9

Lloyd, E.A., Oreskes, N., 2018. Climate Change Attribution: When Is It Appropriate to Accept New Methods? Earth's Future 6, 311–325. https://doi.org/10.1002/2017EF000665

Lo, Y.T.E., Mitchell, D.M., 2021. How will climate change affect UK heatwaves? Weather 76, 326–327. https://doi.org/10.1002/wea.4061

Lo, Y.T.E., Mitchell, D.M., Watson, P.A.G., Screen, J.A., 2023. Changes in Winter Temperature Extremes From Future Arctic Sea-Ice Loss and Ocean Warming. Geophysical Research Letters 50, e2022GL102542. https://doi.org/10.1029/2022GL102542

Lyddon, C., Robins, P., Lewis, M., Barkwith, A., Vasilopoulos, G., Haigh, I., Coulthard, T., 2023. Historic Spatial Patterns of Storm-Driven Compound Events in UK Estuaries. Estuaries and Coasts 46, 30–56. https://doi.org/10.1007/s12237-022-01115-4

Manning, C., Kendon, E.J., Fowler, H.J., Roberts, N.M., 2023. Projected increase in windstorm severity and contribution from sting jets over the UK and Ireland. Weather and Climate Extremes 40, 100562. https://doi.org/10.1016/j.wace.2023.100562

Manning, C., Kendon, E.J., Fowler, H.J., Roberts, N.M., Berthou, S., Suri, D., Roberts, M.J., 2022. Extreme windstorms and sting jets in convection-permitting climate simulations over Europe. Clim Dyn 58, 2387–2404. https://doi.org/10.1007/s00382-021-06011-4

Manning, C., Widmann, M., Bevacqua, E., Loon, A.F.V., Maraun, D., Vrac, M., 2019. Increased probability of compound long-duration dry and hot events in Europe during summer (1950–2013). Environ. Res. Lett. 14, 094006. https://doi.org/10.1088/1748-9326/ab23bf

Martínez-Alvarado, O., Gray, S.L., Hart, N.C.G., Clark, P.A., Hodges, K., Roberts, M.J., 2018. Increased wind risk from sting-jet windstorms with climate change. Environ. Res. Lett. 13, 044002. https://doi.org/10.1088/1748-9326/aaae3a

Massey, N., Aina, T., Rye, C., Otto, F.E.L., Wilson, S., Jones, R.G., Allen, M.R., 2012. Have the odds of warm November temperatures and of cold December temperatures in Central England changed. Bull Am Meteorol Soc 93, 1057–1059.
McCarthy, M., Christidis, N., Dunstone, N., Fereday, D., Kay, G., Klein-Tank, A., Lowe, J., Petch, J., Scaife, A., Stott, P., 2019. Drivers of the UK summer heatwave of 2018. Weather 74, 390–396. https://doi.org/10.1002/wea.3628
McCarthy, M., Christidis, N., Stott, P., Kaye, N., 2021. Met Office: A review of the UK's climate in 2020 [WWW Document]. Carbon Brief. URL https://www.carbonbrief.org/met-office-a-review-of-the-uks-climate-in-2020/ (accessed 6.2.23).
Met Office, 2022. 2022 provisionally warmest year on record for UK [WWW Document]. Met Office. URL https://www.metoffice.gov.uk/about-us/press-office/news/weather-and-climate/2022/2022-provisionally-warmest-year-on-record-for-uk (accessed 6.2.23).
Met Office, 2019a. UK and Global extreme events – Cold [WWW Document]. Met Office. URL https://www.metoffice.gov.uk/research/climate/understanding-climate/uk-and-global-extreme-events-cold (accessed 6.2.23).
Met Office, 2019b. What is a sting jet? [WWW Document]. Met Office. URL https://www.metoffice.gov.uk/weather/learn-about/weather/types-of-weather/wind/sting-jet (accessed 6.2.23).
Mitchell, D., 2021. Climate attribution of heat mortality. Nat. Clim. Chang. 11, 467–468. https://doi.org/10.1038/s41558-021-01049-y
Mitchell, D., Davini, P., Harvey, B., Massey, N., Haustein, K., Woollings, T., Jones, R., Otto, F., Guillod, B., Sparrow, S., Wallom, D., Allen, M., 2017. Assessing mid-latitude dynamics in extreme event attribution systems. Clim Dyn 48, 3889–3901. https://doi.org/10.1007/s00382-016-3308-z
Mitchell, D., Hawker, L., Savage, J., Bingham, R., Lord, N.S., Khan, M.J.U., Bates, P., Durand, F., Hassan, A., Huq, S., Islam, A.S., Krien, Y., Neal, J., Sampson, C., Smith, A., Testut, L., 2022. Increased population exposure to Amphan-scale cyclones under future climates. Climate Resilience and Sustainability 1, e36. https://doi.org/10.1002/cli2.36
Mitchell, D., Heaviside, C., Vardoulakis, S., Huntingford, C., Masato, G., Guillod, B.P., Frumhoff, P., Bowery, A., Wallom, D., Allen, M., 2016. Attributing human mortality during extreme heat waves to anthropogenic climate change. Environ. Res. Lett. 11, 074006. https://doi.org/10.1088/1748-9326/11/7/074006
Mitchell, D.M., 2016. Attributing the forced components of observed stratospheric temperature variability to external drivers. Quarterly Journal of the Royal Meteorological Society 142, 1041–1047. https://doi.org/10.1002/qj.2707
Murray, V., Abrahams, J., Adballah, C., Ahmed, K., Angeles, L., Benouar, D., Brenes, T.A., Chan Hun, C., Cox, S., Douris, J., Fagan, L., Fra Paleo, U., Han, Q., Handmer, J., Hodson, S., Khim, W., Mayner, L., Moody, N., Moraes, L.L., Osvaldo, N.M., Norris, J., Peduzzi, P., Perwaiz, A., Peters, K., Radisch, J., Reichstein, M., Schneider, J., Smith, A., Souch, C., Stevance, A.S., Triyanti, Annisa, Weir, M., Wright, N., Environmental Governance, Environmental Governance, 2021. Hazard Information Profiles: Supplement to: UNDRR-ISC Hazard Definition & Classification Review-Technical Report. https://doi.org/10.24948/2021.05
NOAA, 2017. What is water quality? [WWW Document]. Florida Keys National Marine Sanctuary. URL https://floridakeys.noaa.gov/ocean/waterquality.html (accessed 6.2.23).
O'Neill, B.C., van Aalst, M., Zaiton Ibrahim, Z., Berrang-Ford, L., Bhadwal, S., Buhaug, H., Diaz, D., Frieler, K., Garschagen, M., Magnan, A.K., Midgley, G., Mirzabaev, A., Thomas, A., Warren, R., Jiang, T., Oppenheimer, M., 2022. Key Risks Across Sectors and Regions, in: Climate Change 2022: Impacts, Adaptation, and Vulnerability. Intergovernmental Panel on Climate Change.
Otto, F.E.L., 2017. Attribution of Weather and Climate Events. Annual Review of Environment and Resources 42, 627–646. https://doi.org/10.1146/annurev-environ-102016-060847
Otto, F.E.L., Jan van Oldenborgh, G., Vautard, R., Schwierz, C., 2017. Record June temperatures in western Europe – World Weather Attribution. URL https://www.worldweatherattribution.org/european-heat-june-2017/ (accessed 6.2.23).
Otto, F.E.L., Wiel, K. van der, Oldenborgh, G.J. van, Philip, S., Kew, S.F., Uhe, P., Cullen, H., 2018. Climate change increases the probability of heavy rains in Northern England/Southern Scotland like those of storm Desmond—a real-time event attribution revisited. Environ. Res. Lett. 13, 024006. https://doi.org/10.1088/1748-9326/aa9663

Pall, P., Aina, T., Stone, D.A., Stott, P.A., Nozawa, T., Hilberts, A.G.J., Lohmann, D., Allen, M.R., 2011. Anthropogenic greenhouse gas contribution to flood risk in England and Wales in autumn 2000. Nature 470, 382–385. https://doi.org/10.1038/nature09762

Perkins-Kirkpatrick, S.E., Stone, D.A., Mitchell, D.M., Rosier, S., King, A.D., Lo, Y.T.E., Pastor-Paz, J., Frame, D., Wehner, M., 2022. On the attribution of the impacts of extreme weather events to anthropogenic climate change. Environ. Res. Lett. 17, 024009. https://doi.org/10.1088/1748-9326/ac44c8

Perlwitz, J., 2019. A tug-of-war over the mid-latitudes. Nat Commun 10, 5578. https://doi.org/10.1038/s41467-019-13714-0

Pierce, D.W., Gleckler, P.J., Barnett, T.P., Santer, B.D., Durack, P.J., 2012. The fingerprint of human-induced changes in the ocean's salinity and temperature fields. Geophysical Research Letters 39. https://doi.org/10.1029/2012GL053389

Pörtner, H.-O., Roberts, D.C., Adams, H., Adler, C., Aldunce, P., Ali, E., Begum, R.A., Betts, R., Kerr, R.B., Biesbroek, R., Birkmann, J., Bowen, K., Castellanos, E., Cissé, G., Constable, A., Cramer, W., Dodman, D., Eriksen, S.H., Fischlin, A., Garschagen, M., Glavovic, B., Gilmore, E., Haasnoot, M., Harper, S., Hasegawa, T., Hayward, B., Hirabayashi, Y., Howden, M., Kalaba, K., Kiessling, W., Lasco, R., Lawrence, J., Lemos, M.F., Lempert, R., Ley, D., Lissner, T., Lluch-Cota, S., Loeschke, S., Lucatello, S., Luo, Y., Mackey, B., Maharaj, S., Mendez, C., Mintenbeck, K., Möller, V., Vale, M.M., Morecroft, M.D., Mukherji, A., Mycoo, M., Mustonen, T., Nalau, J., Okem, A., Ometto, J.P., Parmesan, C., Pelling, M., Pinho, P., Poloczanska, E., Racault, M.-F., Reckien, D., Pereira, J., Revi, A., Rose, S., Sanchez-Rodriguez, R., Schipper, E.L.F., Schmidt, D., Schoeman, D., Shaw, R., Singh, C., Solecki, W., Stringer, L., Thomas, A., Totin, E., Trisos, C., van Aalst, M., Viner, D., Wairiu, M., Warren, R., Yanda, P., Ibrahim, Z.Z., Adrian, R., Craig, M., Degvold, F., Ebi, K.L., Frieler, K., Jamshed, A., McMillan, J., Mechler, R., New, M., Simpson, N.P., Stevens, N., 2022. Climate Change 2022: Impacts, Adaptation, and Vulnerability. Contribution of Working Group II to the Sixth Assessment Report of the Intergovernmental Panel on Climate Change. Summary for Policymakers.

Priestley, M.D.K., Catto, J.L., 2022. Future changes in the extratropical storm tracks and cyclone intensity, wind speed, and structure. Weather and Climate Dynamics 3, 337–360. https://doi.org/10.5194/wcd-3-337-2022

Ranasinghe, R., Ruane, A.C., Vautard, R., Arnell, N., Coppola, E., Cruz, F.A., Dessai, S., Saiful Islam, A.K.M., Rahimi, M., Carrascal, D.R., Sillmann, J., Sylla, M.B., Tebaldi, C., Wang, W., Zaaboul, R., 2021. Climate change information for regional impact and for risk assessment, in: Climate Change 2021: The Physical Science Basis. Contribution of Working Group I to the Sixth Assessment Report of the Intergovernmental Panel on Climate Change. Cambridge University Press, Cambridge, pp. 1767–1926.

Raupach, T.H., Martius, O., Allen, J.T., Kunz, M., Lasher-Trapp, S., Mohr, S., Rasmussen, K.L., Trapp, R.J., Zhang, Q., 2021. The effects of climate change on hailstorms. Nat Rev Earth Environ 2, 213–226. https://doi.org/10.1038/s43017-020-00133-9

Rhein, M., Rintoul, S.R., Aoki, S., Campos, E., Chambers, D., Feely, R.A., Gulev, S., Johnson, G.C., Josey, S.A., Kostianoy, A., 2013. Observations: Ocean in Climate Change 2013: The Physical Science Basis. Contribution of Working Group I to the Fifth Assessment Report of the Intergovernmental Panel on Climate Change. Fifth assessment report of the Intergovernmental Panel on Climate Change 255–316.

Sabine, C.L., Feely, R.A., Gruber, N., Key, R.M., Lee, K., Bullister, J.L., Wanninkhof, R., Wong, C.S., Wallace, D.W.R., Tilbrook, B., Millero, F.J., Peng, T.-H., Kozyr, A., Ono, T., Rios, A.F., 2004. The Oceanic Sink for Anthropogenic CO2. Science 305, 367–371. https://doi.org/10.1126/science.1097403

Sahani, J., Kumar, P., Debele, S., Emmanuel, R., 2022. Heat risk of mortality in two different regions of the United Kingdom. Sustainable Cities and Society 80, 103758. https://doi.org/10.1016/j.scs.2022.103758

Sanderson, M.G., Hand, W.H., Groenemeijer, P., Boorman, P.M., Webb, J.D.C., McColl, L.J., 2015. Projected changes in hailstorms during the 21st century over the UK. International Journal of Climatology 35, 15–24. https://doi.org/10.1002/joc.3958

Schaller, N., Kay, A.L., Lamb, R., Massey, N.R., van Oldenborgh, G.J., Otto, F.E.L., Sparrow, S.N., Vautard, R., Yiou, P., Ashpole, I., Bowery, A., Crooks, S.M., Haustein, K., Huntingford, C., Ingram, W.J., Jones, R.G., Legg, T., Miller, J., Skeggs, J., Wallom, D., Weisheimer, A., Wilson, S., Stott, P.A., Allen, M.R., 2016. Human influence on climate in the 2014 southern England winter floods and their impacts. Nature Clim Change 6, 627–634. https://doi.org/10.1038/nclimate2927

Slangen, A.B.A., Church, J.A., Zhang, X., Monselesan, D., 2014. Detection and attribution of global mean thermosteric sea level change. Geophysical Research Letters 41, 5951–5959. https://doi.org/10.1002/2014GL061356

Slangen, A.B.A., Church, J.A., Zhang, X., Monselesan, D.P., 2015. The Sea Level Response to External Forcings in Historical Simulations of CMIP5 Climate Models. Journal of Climate 28, 8521–8539. https://doi.org/10.1175/JCLI-D-15-0376.1

Slingo, J., 2021. Latest Scientific Evidence for Observed and Projected Climate Change. Technical Report of the Third UK Climate Change Risk Assessment, Technical Report of the Third UKL Climate Change Risk Assessment Chapter 1.

Sparrow, S., Huntingford, C., Massey, N., Allen, M.R., 2013. 12. THE USE OF A VERY LARGE ATMOSPHERIC MODEL ENSEMBLE TO ASSESS POTENTIAL ANTHROPOGENIC INFLUENCE ON THE UK SUMMER 2012 HIGH RAINFALL TOTALS.

Stone, D., Auffhammer, M., Carey, M., Hansen, G., Huggel, C., Cramer, W., Lobell, D., Molau, U., Solow, A., Tibig, L., Yohe, G., 2013. The challenge to detect and attribute effects of climate change on human and natural systems. Climatic Change 121, 381–395. https://doi.org/10.1007/s10584-013-0873-6

Stone, D.A., Hansen, G., 2016. Rapid systematic assessment of the detection and attribution of regional anthropogenic climate change. Clim Dyn 47, 1399–1415. https://doi.org/10.1007/s00382-015-2909-2

Stott, P.A., Christidis, N., 2023. Operational attribution of weather and climate extremes: what next? Environ. Res.: Climate 2, 013001. https://doi.org/10.1088/2752-5295/acb078

Stott, P.A., Christidis, N., Otto, F.E.L., Sun, Y., Vanderlinden, J.-P., van Oldenborgh, G.J., Vautard, R., von Storch, H., Walton, P., Yiou, P., Zwiers, F.W., 2016. Attribution of extreme weather and climate-related events. WIREs Climate Change 7, 23–41. https://doi.org/10.1002/wcc.380

Stott, P.A., Stone, D.A., Allen, M.R., 2004. Human contribution to the European heatwave of 2003. Nature 432, 610–614. https://doi.org/10.1038/nature03089

Stuart-Smith, R., Saad, A., Otto, F., Lisi, G., Lauta, K., Wetzer, T., 2021. Attribution science and litigation: facilitating effective legal arguments and strategies to manage climate change damages. FILE Foundation.

Surminski, S., Baglee, A., Cameron, C., Connell, R., Deyes, K., Haworth, A., Ingirige, B., Muir-Wood, R., Proverbs, D., Watkiss, P., 2021. Business and industry. The Third UK Climate Change Risk Assessment Technical Report 1–189.

Trenberth, K.E., 2011. Changes in precipitation with climate change. Climate Research 47, 123–138. https://doi.org/10.3354/cr00953

Uhe, P., Mitchell, D., Bates, P.D., Addor, N., Neal, J., Beck, H.E., 2021a. Model cascade from meteorological drivers to river flood hazard: flood-cascade v1.0. Geoscientific Model Development 14, 4865–4890. https://doi.org/10.5194/gmd-14-4865-2021

Uhe, P., Mitchell, D., Bates, P.D., Allen, M.R., Betts, R.A., Huntingford, C., King, A.D., Sanderson, B.M., Shiogama, H., 2021b. Method Uncertainty Is Essential for Reliable Confidence Statements of Precipitation Projections. Journal of Climate 34, 1227–1240. https://doi.org/10.1175/JCLI-D-20-0289.1

Uhe, P., Otto, F.E.L., Haustein, K., van Oldenborgh, G.J., King, A.D., Wallom, D.C.H., Allen, M.R., Cullen, H., 2016. Comparison of methods: Attributing the 2014 record European temperatures to human influences. Geophysical Research Letters 43, 8685–8693. https://doi.org/10.1002/2016GL069568

Undorf, S., Allen, K., Hagg, J., Li, S., Lott, F.C., Metzger, M.J., Sparrow, S.N., Tett, S.F.B., 2020. Learning from the 2018 heatwave in the context of climate change: are high-temperature extremes important for adaptation in Scotland? Environ. Res. Lett. 15, 034051. https://doi.org/10.1088/1748-9326/ab6999

USGS, 2021. What is a landslide and what causes one? | U.S. Geological Survey [WWW Document]. URL https://www.usgs.gov/faqs/what-landslide-and-what-causes-one (accessed 6.2.23).

Vautard, R., Aalst, M. van, Boucher, O., Drouin, A., Haustein, K., Kreienkamp, F., Oldenborgh, G.J. van, Otto, F.E.L., Ribes, A., Robin, Y., Schneider, M., Soubeyroux, J.-M., Stott, P., Seneviratne, S.I., Vogel, M.M., Wehner, M., 2020. Human contribution to the record-breaking June and July 2019 heatwaves in Western Europe. Environ. Res. Lett. 15, 094077. https://doi.org/10.1088/1748-9326/aba3d4

Vautard, R., Yiou, P., Otto, F., Stott, P., Christidis, N., Oldenborgh, G.J. van, Schaller, N., 2016. Attribution of human-induced dynamical and thermodynamical contributions in extreme weather events. Environ. Res. Lett. 11, 114009. https://doi.org/10.1088/1748-9326/11/11/114009

Vicedo-Cabrera, A.M., Scovronick, N., Sera, F., Royé, D., Schneider, R., Tobias, A., Astrom, C., Guo, Y., Honda, Y., Hondula, D.M., Abrutzky, R., Tong, S., Coelho, M. de S.Z.S., Saldiva, P.H.N., Lavigne, E., Correa, P.M., Ortega, N.V., Kan, H., Osorio, S., Kyselý, J., Urban, A., Orru, H., Indermitte, E., Jaakkola, J.J.K., Ryti, N., Pascal, M., Schneider, A., Katsouyanni, K., Samoli, E., Mayvaneh, F., Entezari, A., Goodman, P., Zeka, A., Michelozzi, P., de'Donato, F., Hashizume, M., Alahmad, B., Diaz, M.H., Valencia, C.D.L.C., Overcenco, A., Houthuijs, D., Ameling, C., Rao, S., Di Ruscio, F., Carrasco-Escobar, G., Seposo, X., Silva, S., Madureira, J., Holobaca, I.H., Fratianni, S., Acquaotta, F., Kim, H., Lee, W., Iniguez, C., Forsberg, B., Ragettli, M.S., Guo, Y.L.L., Chen, B.Y., Li, S., Armstrong, B., Aleman, A., Zanobetti, A., Schwartz, J., Dang, T.N., Dung, D.V., Gillett, N., Haines, A., Mengel, M., Huber, V., Gasparrini, A., 2021. The burden of heat-related mortality attributable to recent human-induced climate change. Nat. Clim. Chang. 11, 492–500. https://doi.org/10.1038/s41558-021-01058-x

Vogel, M.M., Zscheischler, J., Wartenburger, R., Dee, D., Seneviratne, S.I., 2019. Concurrent 2018 Hot Extremes Across Northern Hemisphere Due to Human-Induced Climate Change. Earth's Future 7, 692–703. https://doi.org/10.1029/2019EF001189

Watkiss, P., Betts, R., 2021. Latest Scientific Evidence for Observed and Projected Climate Change. Technical Report of the Third UK Climate Change Risk Assessment, Technical Report of the Third UKL Climate Change Risk Assessment Chapter 2.

Wehrli, K., Guillod, B.P., Hauser, M., Leclair, M., Seneviratne, S.I., 2018. Assessing the Dynamic Versus Thermodynamic Origin of Climate Model Biases. Geophysical Research Letters 45, 8471–8479. https://doi.org/10.1029/2018GL079220

Werritty, A., 2002. Living with uncertainty: climate change, river flows and water resource management in Scotland. Science of The Total Environment, Hydrology in Scotland: towards a scientific basis for the sustainable management of freshwater resources 294, 29–40. https://doi.org/10.1016/S0048-9697(02)00050-5

WHO, 2022. Air pollution [WWW Document]. URL https://www.who.int/health-topics/air-pollution (accessed 6.2.23).

Wilcox, L.J., Yiou, P., Hauser, M., Lott, F.C., van Oldenborgh, G.J., Colfescu, I., Dong, B., Hegerl, G., Shaffrey, L., Sutton, R., 2018. Multiple perspectives on the attribution of the extreme European summer of 2012 to climate change. Clim Dyn 50, 3537–3555. https://doi.org/10.1007/s00382-017-3822-7

Wild, S., Befort, D.J., Leckebusch, G.C., 2015. 7. Was the Extreme Storm Season in Winter 2013/14 Over the North Atlantic and the United Kingdom Triggered by Changes in the West Pacific Warm Pool? Bulletin of the American Meteorological Society 96, S29–S34.

WMO, 2020. Thunderstorm [WWW Document]. International Cloud Atlas. URL https://cloudatlas.wmo.int/thunderstorm.html (accessed 6.2.23).

World Weather Attribution, 2014. 2014 likely to be the warmest year ever recorded – World Weather Attribution. URL https://www.worldweatherattribution.org/european-heat-2014/ (accessed 6.2.23).

Yule, E.L., Hegerl, G., Schurer, A., Hawkins, E., 2023. Using early extremes to place the 2022 UK heat waves into historical context. Atmospheric Science Letters n/a, e1159. https://doi.org/10.1002/asl.1159

Zachariah, M., Vautard, R., Schumacher, D.L., 2022. Without human-caused climate change temperatures of 40°C in the UK would have been extremely unlikely – World Weather Attribution. URL https://www.worldweatherattribution.org/without-human-caused-climate-change-temperatures-of-40c-in-the-uk-would-have-been-extremely-unlikely/ (accessed 6.2.23).

Zhai, P., Zhou, B., Chen, Y., 2018. A Review of Climate Change Attribution Studies. J Meteorol Res 32, 671–692. https://doi.org/10.1007/s13351-018-8041-6

Zhang, X., Zwiers, F.W., Hegerl, G.C., Lambert, F.H., Gillett, N.P., Solomon, S., Stott, P.A., Nozawa, T., 2007. Detection of human influence on twentieth-century precipitation trends. Nature 448, 461–465. https://doi.org/10.1038/nature06025

Zurich, 2022. Three common types of floods explained [WWW Document]. URL https://www.zurich.com/en/knowledge/topics/flood-and-water-damage/three-common-types-of-flood (accessed 6.2.23).